\documentclass[twoside,twocolumn,9pt]{article}
\usepackage{extsizes}
\usepackage[super,sort&compress,comma]{natbib} 
\usepackage[version=3]{mhchem}   
\usepackage[left=1.5cm, right=1.5cm, top=1.785cm, bottom=2.0cm]{geometry}
\usepackage{balance}
\usepackage{times,mathptmx}
\usepackage{sectsty}
\usepackage{graphicx} 
\usepackage{lastpage}
\usepackage[format=plain,justification=justified,singlelinecheck=false,font={stretch=1.125,small,sf},labelfont=bf,labelsep=space]{caption}
\usepackage{float}
\usepackage{fancyhdr}
\usepackage{fnpos}

\usepackage[english]{babel}
\usepackage{amssymb}
\addto{\captionsenglish}{%
  
}  
\usepackage{array}
\usepackage{droidsans}
\usepackage{charter}
\usepackage[T1]{fontenc}
\usepackage[usenames,dvipsnames]{xcolor}
\usepackage{setspace}
\usepackage[compact]{titlesec}
\usepackage{hyperref}

\usepackage{epstopdf}

\def\unit#1{\ensuremath{\mathrm{\,#1}}}
\def\celsius{\ensuremath{\mathrm{^{\circ}C}}}
\def\E{\ensuremath{\mathrm{e}}}
\def\Fbond{\ensuremath{\mathcal{F}_{\text{bond}}}}
\def\NbNL{\ensuremath{\#_b^{\text{NL}}}}
\def\NNL{\ensuremath{N_{\text{NL}}}}
\def\NcNL{\ensuremath{N_c^{\text{NL}}}}
\def\QnNL{\ensuremath{Q_n^{\text{NL}}}}

\def\NcL{\ensuremath{N_c^{\text{L}}}}
\def\QnL{\ensuremath{Q_n^{\text{L}}}}
\def\pbFS{\ensuremath{p_b^{\text{FS}}}}

\definecolor{cream}{RGB}{222,217,201}

\usepackage{pdfpages}

\begin{document}

\pagestyle{fancy}
\thispagestyle{plain}
\fancypagestyle{plain}{

}

\makeFNbottom
\makeatletter
\renewcommand\LARGE{\@setfontsize\LARGE{15pt}{17}}
\renewcommand\Large{\@setfontsize\Large{12pt}{14}}
\renewcommand\large{\@setfontsize\large{10pt}{12}}
\renewcommand\footnotesize{\@setfontsize\footnotesize{7pt}{10}}
\makeatother

\renewcommand{\thefootnote}{\fnsymbol{footnote}}
\renewcommand\footnoterule{\vspace*{1pt}%
\color{cream}\hrule width 3.5in height 0.4pt \color{black}\vspace*{5pt}} 
\setcounter{secnumdepth}{5}

\makeatletter 
\renewcommand\@biblabel[1]{#1}            
\renewcommand\@makefntext[1]%
{\noindent\makebox[0pt][r]{\@thefnmark\,}#1}
\makeatother 
\renewcommand{\figurename}{\small{Fig.}~}
\sectionfont{\sffamily\Large}
\subsectionfont{\normalsize}
\subsubsectionfont{\bf}
\setstretch{1.125} 
\setlength{\skip\footins}{0.8cm}
\setlength{\footnotesep}{0.25cm}
\setlength{\jot}{10pt}
\titlespacing*{\section}{0pt}{4pt}{4pt}
\titlespacing*{\subsection}{0pt}{15pt}{1pt}

\fancyfoot{}
\fancyfoot[LO,RE]{\vspace{-7.1pt}\includegraphics[height=9pt]{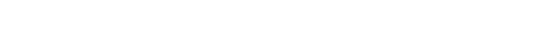}}
\fancyfoot[CO]{\vspace{-7.1pt}\hspace{13.2cm}\includegraphics{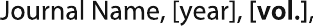}}
\fancyfoot[CE]{\vspace{-7.2pt}\hspace{-14.2cm}\includegraphics{head_foot/RF}}
\fancyfoot[RO]{\footnotesize{\sffamily{1--\pageref{LastPage} ~\textbar  \hspace{2pt}\thepage}}}
\fancyfoot[LE]{\footnotesize{\sffamily{\thepage~\textbar\hspace{3.45cm} 1--\pageref{LastPage}}}}
\fancyhead{}
\renewcommand{\headrulewidth}{0pt} 
\renewcommand{\footrulewidth}{0pt}
\setlength{\arrayrulewidth}{1pt}
\setlength{\columnsep}{6.5mm}
\setlength\bibsep{1pt}

\makeatletter 
\newlength{\figrulesep} 
\setlength{\figrulesep}{0.5\textfloatsep} 

\newcommand{\topfigrule}{\vspace*{-1pt}%
\noindent{\color{cream}\rule[-\figrulesep]{\columnwidth}{1.5pt}} }

\newcommand{\botfigrule}{\vspace*{-2pt}%
\noindent{\color{cream}\rule[\figrulesep]{\columnwidth}{1.5pt}} }

\newcommand{\dblfigrule}{\vspace*{-1pt}%
\noindent{\color{cream}\rule[-\figrulesep]{\textwidth}{1.5pt}} }

\makeatother

\twocolumn[
  \begin{@twocolumnfalse}
  {\includegraphics[height=30pt]{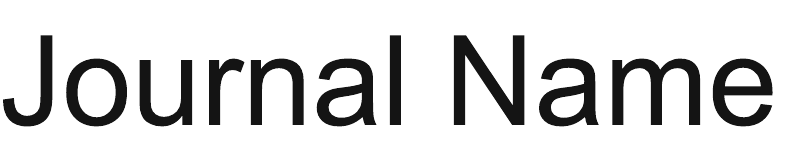}\hfill%
 \raisebox{0pt}[0pt][0pt]{\includegraphics[height=55pt]{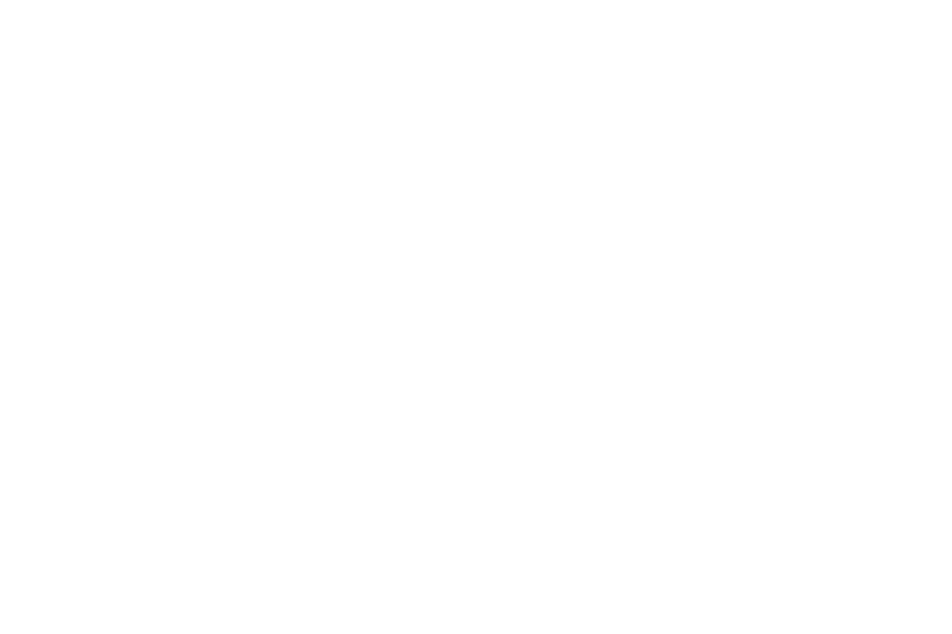}}%
 \\[1ex]%
 \includegraphics[width=18.5cm]{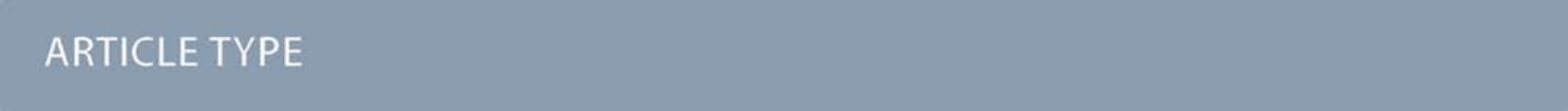}}\par
\vspace{1em}
\sffamily
\begin{tabular}{m{4.5cm} p{13.5cm} }

\includegraphics{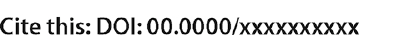} & \noindent\LARGE{\textbf{Hyperbranched DNA clusters$^\dag$}} \\
\vspace{0.3cm} & \vspace{0.3cm} \\

 & \noindent\large{Enrico Lattuada,\textit{$^{a}$} Debora Caprara,\textit{$^{a}$} Vincenzo Lamberti,\textit{$^{a}$} and Francesco Sciortino\textit{$^{a}$}} \\

\includegraphics{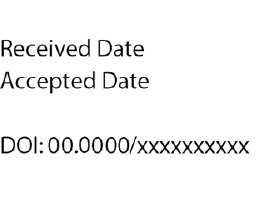} & \noindent\normalsize{Taking advantage of the base-pairing specificity and tunability of DNA interactions, we investigate the spontaneous formation of hyperbranched clusters starting from purposely designed DNA tetravalent nanostar monomers, encoding in their four sticky-ends the desired binding rules.
Specifically, we combine molecular dynamics simulations  and Dynamic Light Scattering experiments to follow the aggregation process of the DNA nanostars at different concentrations and temperatures. 
At odd with the Flory-Stockmayer predictions, we find that, even when all possible bonds are formed, the system does not reach percolation due to the presence of intracluster bonds.
We present an extension of the Flory-Stockmayer theory that properly describes the numerical and the experimental results.
} \\

\end{tabular}

 \end{@twocolumnfalse} \vspace{0.6cm}

 ]

\renewcommand*\rmdefault{bch}\normalfont\upshape
\rmfamily
\section*{}
\vspace{-1cm}


\footnotetext{\textit{$^{a}$~Physics Department, Sapienza University, P.le Aldo Moro 5, 00185, Rome, Italy.}}

\footnotetext{\dag~Electronic Supplementary Information (ESI) available: [details of any supplementary information available should be included here]. See DOI: 00.0000/00000000.}



In recent years, the relation between reversible self-assembly of patchy colloidal particles and irreversible aggregation of chemical units is receiving a renewed interest~\cite{Bianchi2007,Corezzi2008}.
This connection has been nourished by the observation that the clusters considered in the Wertheim theory for associating liquids~\cite{Wertheim1984,Wertheim1986} are the same loopless clusters considered in the Flory-Stockmayer (FS) theory of polyfunctional condensation~\cite{Flory1941,Stockmayer1943}.
For example, the  conditions for the formation of infinite networks, valid for describing the self-condensation of $f$-functional $A_f$ monomers, are identical to  the ones for identical colloids with $f$ patches. In contrast to the chemical case, for which equilibrium conditions are assumed -- but hardly realizable in experiments due to the covalent nature of the bonds --, colloidal aggregation may proceed to equilibrium, allowing for a more accurate control of the theoretical FS predictions.

For the case of identical colloids with $f$ patches, there are evidences that the range of validity of the FS predictions becomes wider and wider on decreasing $f$~\cite{Bianchi2006,Sciortino2007}.
For binary mixtures of very small average ``valence'' $\langle f \rangle$, the FS predictions provide a quite accurate description of the cluster size distribution, except for bond probabilities very close to the percolation threshold (itself properly predicted theoretically)~\cite{Bianchi2007}.
For this reason, patchy colloidal particles have become a test ground for revisiting the old FS predictions in proper equilibrium conditions.
Furthermore, they provided a way to access the role of the bonding loops (\emph{i.e.} close paths of bonds), which are commonly neglected both in FS and Wertheim theories.

An interesting aggregation phenomenon takes place in one-component systems made of monomers of $AB_{f-1}$ type\textit{$^{*}$}\footnotetext{\textit{$^{*}$~The $AB_{f-1}$ monomer can also be indicated as $A-R-B_{f-1}$ or $ARB_{f-1}$.}}, where the $A$ site condenses with $B$ but reactions between like functional groups ($AA$ and $BB$) are forbidden.
The clusters resulting from this aggregation process are commonly known as \emph{hyperbranched} polymers, a term introduced by Kim and Webster in their works on the synthesis of highly branched polymers~\cite{Kim1990,Kim1992}.
In the last decades, the interest towards the synthesis and understanding of these materials has continuously grown, representing a challenge for innovative applications.
Hyperbranched polymers constitute an appealing alternative to dendrimers, owing to their facile synthesis and high tunability~\cite{Cuneo2020,Jochum2019}.
Similarly to other branched polymers, they are characterized by high exposure of functional groups, three-dimensional globular structure, low viscosity, and good solubility~\cite{Zhou2010,Liu2015}.
Potential applications include surface coating~\cite{vanBenthem2000}, use as filler in composite materials to increase the thermal and mechanical stability~\cite{Zotti2019}, drug and gene delivery~\cite{Paleos2010,Gajbhiye2013,Qi2016,Zhao2017,Song2019}, grafting on nanoparticles for diagnostic imaging to reduce the toxicity~\cite{Feng2012,MashhadiMalekzadeh2017}, and sensors~\cite{Pitois2001,Shi2013,Liu2014}.

From the theoretical standpoint, the aggregation of $AB_{f-1}$ units is particularly interesting for several reasons:
(i) it is analytically tractable (neglecting the formation of closed bond loops);
(ii) the cluster size distributions for branched polymers are requested as
an intermediate step  in the evaluation of several 
polyfunctional condensation processes;
(iii) it gives rise, according to the FS theory,  to an aggregation phenomenon in which the fully bonded case (when all $A$ sites have reacted) corresponds to the percolation transition.
Therefore, hyperbranched polymers do not have a gel phase, but only a sol one.

Despite its interest, an accurate comparison between FS theoretical predictions and  numerical and experimental results in the case of hyperbrached polymers has rarely been attempted~\cite{Zheng2011,Lyu2018}.
Actually, the hypothesis of absence of intramolecular reactions is expected to get progressively worse for large degrees of polymerization.
In fact, an unreacted $A$ will most likely interact with one of the (nearby) $B$ sites belonging to the same cluster, hence forming closed loops.
As a result, the cluster size distribution of hyperbranched polymers may not follow the FS predictions.

In line with the conceptual framework of colloids/reacting monomers, we present here a combined numerical and experimental study of a colloidal analog of the $AB_{f-1}$ hyperbranching condensation.
In particular, we design specific DNA oligomers able to self-assemble into bulk quantities of identical four-armed particles which can interact in a controlled way~\cite{Biffi2013,Biffi2015,Rovigatti2014,Conrad2019}.
We exploit molecular dynamics (MD) simulations based on the oxDNA2 coarse-grained interaction potential~\cite{Ouldridge2011,Snodin2015} to follow the particle aggregation process and to compare the numerical results with theoretical predictions.
This numerical study allows us to estimate the role of intramolecular binding and how the presence of closed loops modifies the cluster size distribution.
We then demonstrate that a cluster-based thermodynamic treatment, which includes also intracluster bonds, can be developed to extend the FS theory in order to accurately describe the simulation data and to provide a significant guide to the experimental results.
Finally, we realize the same system in the laboratory and follow the aggregation process via Dynamic Light Scattering (DLS) to provide evidence that the presence of intracluster bonds prevents the formation of a percolating state, even in the limit of full bonding.

\section*{The DNA particle}
\begin{figure}[tb]
\includegraphics[width=\columnwidth]{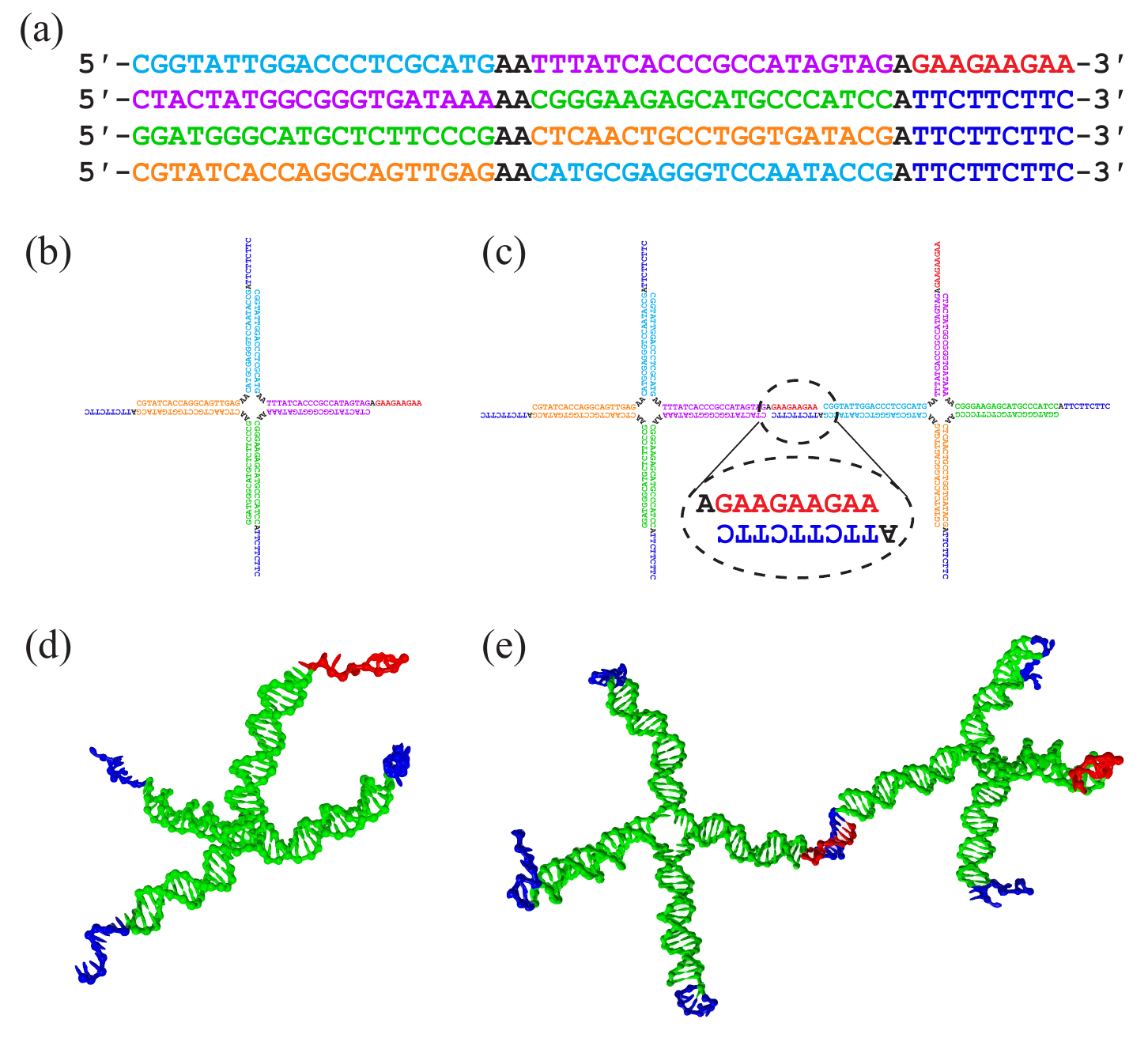}
\caption{(a) Oligonucleotide composition of the strands comprising the tetravalent monomer. Colors correspond to sequence pairings forming the double-stranded monomer arms, shown in (b). The last 9 bases are the $A$ (red) and $B$ (blue) sticky-end sequences, respectively. (c) NS-NS binding via the 9-base long   DNA sequences, located at the  tips. (d-e) Corresponding images generated from oxDNA configurations.}
\label{fig:schematic_monomers}
\end{figure}

In the last years, the ideas born with DNA nanotechnology~\cite{Seeman2003,Seeman2005} have been borrowed by the soft matter community to synthesize bulk quantities of nanometric particles with controlled shape and binding properties~\cite{Bellini2011,Biffi2013,Biffi2015,Conrad2019,Salamonczyk2016}, including dendrimers~\cite{Jochum2019}.
The particles we envision in this study are assembled starting from four distinct DNA oligomers, each composed by 52 nucleotides, containing properly designed sequences of complementary groups (see Fig.~\ref{fig:schematic_monomers}a).
The self-assembly of these strands generates a well-defined nanostar (NS), composed of four double-stranded arms of 20 base pairs departing from a flexible core of 8 unpaired adenines, which provides flexibility to the particles (Fig.~\ref{fig:schematic_monomers}b,d).
Each arm terminates with a 9-base long single-stranded sticky sequence preceded by an additional unbonded adenine, which is inserted to ease the linking between different NSs.

Most of previous works on DNA NSs focused on the $A_4$ case, tetra-armed NSs with identical self-complementary sticky sequences, originating all identical $AA$ bonds~\cite{Biffi2013,Biffi2015,Rovigatti2014,FernandezCastanon2016,Nguyen2017,Bomboi2015}.
The $A_4$ system displays the analog of the gas-liquid phase separation at low temperatures, driven by the association of the self-complementary sticky-ends.
For concentrations larger than the ``liquid'' coexistence value, the system forms a reversible equilibrium gel, which exhibits the static and dynamic features expected for colloidal particles of valence four~\cite{Biffi2013,Biffi2015}.

With the specific design here presented, an individual NS possesses one sticky-end of type $A$ and three of type $B$ in order to mimic a tetravalent monomer of the form $AB_3$.
The $A$ and $B$ sticky regions have been adequately selected to allow only $AB$ bonds between different NSs (Fig.~\ref{fig:schematic_monomers}c,e).

The temperature response of the system is strictly related to the number of nucleotides in the complementary sequences, whose length allows to distinguish different hierarchical self-assembly processes.
As reported in Fig.~S1 of the ESI\dag, above $T = T_{\text{NS}} \approx 77\celsius$ the sample is composed of single and freely-diffusing DNA strands.
Around $T_{\text{NS}}$, the complementary sequences comprising the double-stranded arms start to self-assemble, giving rise to a solution of unbonded NSs.
On further cooling, around $T = T_b \approx 42\celsius$, the sticky-ends start to pair forming inter-, and possibly intra-, NSs $AB$ bonds.
At ambient  temperature and below, essentially all possible bonds are formed.

\section*{Materials and methods}
\subsection*{Numerical methods}

To simulate the aggregation kinetics of $AB_3$ DNA NSs, we employ the coarse-grained model oxDNA2, which is able to reproduce the structural and thermodynamic properties of single- and double-stranded DNA molecules~\cite{Ouldridge2011,Snodin2015}.
The interactions between nucleotides account for excluded volume, backbone connectivity, Watson-Crick hydrogen bonding, stacking, cross-stacking, coaxial-stacking, and for electrostatic interactions at salt concentrations $c_{\text{Na}^+}>0.1\unit{M}$.
The model parameters have been adjusted to reproduce the experimental melting temperature data~\cite{SantaLucia1998,Holbrook1999,Ouldridge2011}.
A code implementing the oxDNA2 model is freely available~\cite{oxdna-code}.

Initial configurations are generated by randomly placing copies of an already assembled DNA tetramer in the simulation box, fulfilling basic steric excluded volume requirements.
The volume is computed for the different values of $N$ and concentrations ($c = 2\unit{mg/ml}$, $10\unit{mg/ml}$, and $20\unit{mg/ml}$, equal to those experimentally investigated), using a tetramer molecular weight of $M_w = 63893\unit{g/mol}$. 
We perform MD simulations in the $NVT$ ensemble with systems consisting of $N=300$, 1000, and 2000 DNA NSs of $AB_3$ type.
The largest system corresponds to $\sim 4 \times 10^5$ interaction sites.
The temperature in the simulation, kept at $T=45\celsius$, is enforced by an Anderson-like thermostat that emulates a Brownian motion~\cite{Russo2009}.
For the selected temperature, we are able to follow the equilibration of the system for up to $10^{10}$ MD timesteps (corresponding to $\sim 30\unit{\mu s}$ of real time and six-months of continuous computation per run), taking advantage of the computational power of modern Graphic Processing Units (GPU).

\subsection*{Experimental methods}
DNA sequences are purchased from Integrated DNA Technologies (IDT) with PAGE purification.
Lyophilized samples are initially dispersed in a filtered, DNAse-free, $50\unit{mM}$ NaCl solution.
Tetravalent particles are pre-assembled by mixing equimolar quantities of the single-stranded components to a final NS concentration $c = 22\unit{mg/ml}$ ($344\unit{\mu M}$).
The mixture is heated up to $90\celsius$, incubated for 20 minutes, and slowly cooled down to room temperature overnight. The NS anneling is carried out using a Memmert oven.

We experimentally investigate three samples at different NS concentrations: $2\unit{mg/ml}$, $10\unit{mg/ml}$, and $20\unit{mg/ml}$.
The samples are prepared into borosilicate glass capillaries (inner diameter $2.4\unit{mm}$, Hilgenberg GmbH).
We dilute the NS batch suspension with a NaCl solution at a proper molarity, in order to obtain $30\unit{\mu l}$ of each sample at a final NaCl concentration of $250\unit{mM}$.
Finally, we cover the suspension with $20\unit{\mu l}$ of silicon oil and seal the capillary using UV resin to avoid sample evaporation.

DLS measurements are carried out at a fixed angle $\theta = 90^{\circ}$ with a custom-made setup consisting of a $633\unit{nm}$ He-Ne Laser ($17\unit{mW}$, Newport Corp.) and a multi-tau digital correlator (Brookhaven Inst.) connected to an optical fiber.
Samples are immersed in a water bath connected to a thermostat.
The actual temperature of the bath near the sample is measured using a thermocouple probe with a $\Delta T = \pm 0.05\celsius$ accuracy.

For each selected temperature, within the interval $10\celsius \le T \le 50\celsius$ every $\Delta T \simeq 5\celsius$, the sample is  thermalized for 40 minutes before starting the acquisition.
Each measurement lasts 10 minutes.
The autocorrelation functions of the scattered intensity $g_2(t)$ are calculated from the correlator output and converted into the field correlation functions $g_1(t)$ using the Siegert relation.

\section*{Results}
\subsection*{Numerical results}
We start by examining the number of bonds $\#_{b}(t)$ as a function of time.
We consider two NSs to be bound when at least 5 complementary bases of the $AB$ sticky sequences hybridize.
Since the maximum number of possible bonds (fully bonded state) is equal to the number of particles $N$, the fraction of bonds $\#_{b}(t) / N$ coincides with the probability $p_b(t)$ that an arbitrary $A$ sticky-end is engaged in a bond with a $B$ overhang at time $t$.
At long times, when equilibrium is reached, $p_b$ approaches the (concentration-dependent) value fixed by the binding equilibrium constant~\cite{SantaLucia1996}.

The FS mean-field theory~\cite{Flory1941,Rubinstein2003} provides a prediction for the number of clusters  $N_c(n)$ of size $n$
\begin{equation}
\label{eq:csd}
N_c (n) = N (1-p_b) F (n,p_b) ,
\end{equation}
where
\begin{equation}
F (n,p_b) = \frac{[(f-1)n]!}{n![(f-2)n+1]!} \frac{p_b^{n-1} (f-1-p_b)^{(f-2)n+1}}{(f-1)^{(f-1)n}}
\label{eq:csd2}
\end{equation}
and the distribution is normalized such that
\begin{equation}
\sum_n n N_c(n) = N .
\end{equation}
The FS theory, which is formally equivalent to a constrained maximization of the combinatorial entropy~\cite{Sciortino2019}, is based on the evaluation of the number of distinct cluster arrangements, with the restriction that the $N$ monomers are connected by $N p_b$ bonds to form polydisperse loopless clusters.
In Eq.~\eqref{eq:csd}, the term $N (1-p_b)$ is equal to the total number of clusters and clearly reveals the mean-field approximation, being the number of clusters equal to the number of particles minus the number of bonds ($N p_b$).

\begin{figure}[tb]
\includegraphics[width=0.5\textwidth]{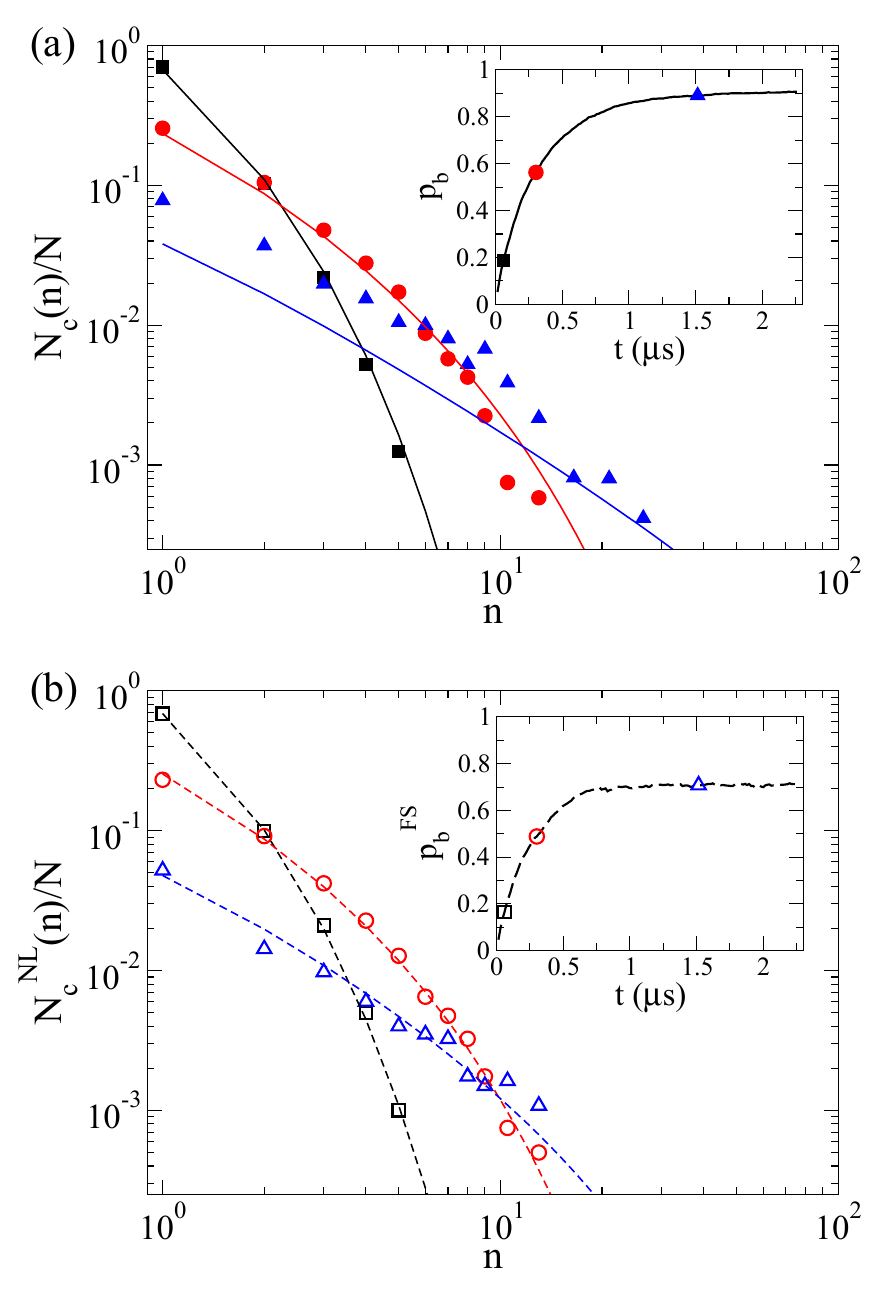}
\caption{The main graphs show the number distribution of clusters (symbols, panel a) and the number distribution of clusters in the FS subset (b) of size $n$ for different simulation times (\emph{i.e.} different bonding probabilities) for the simulation at $c=20\unit{mg/ml}$, $N=2000$. The lines are the theoretical predictions given by Eq.~\eqref{eq:csd} using $p_b$ (panel a) and $\pbFS$ (panel b), respectively, whose evolution over the simulation time is displayed in the insets (symbols and colors correspond to the timestep relative to the curves in the main graphs).
}
\label{fig:csd_20mgml}
\end{figure}

\begin{figure}[tb]
\includegraphics[width=0.5\textwidth]{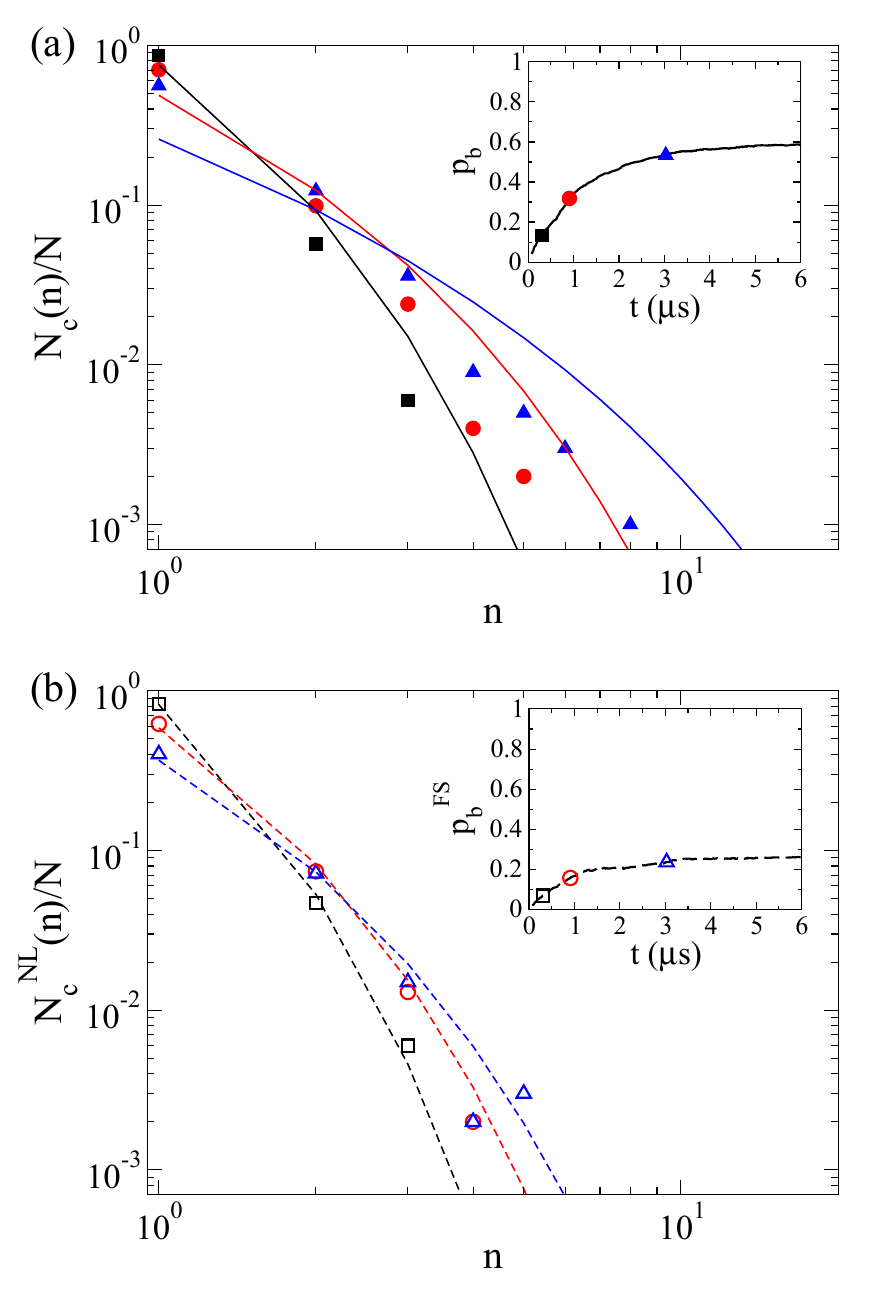}
\caption{Same quantities as the ones presented in Fig.~\ref{fig:csd_20mgml}, but for the simulation at $c=2\unit{mg/ml}$, $N=1000$.}
\label{fig:csd_2mgml}
\end{figure}

Figs.~\ref{fig:csd_20mgml}a and~\ref{fig:csd_2mgml}a show the cluster size distribution $N_c(n)/N$ at three different values of $p_b$ (three different times during the simulation) for two different NS concentrations ($c=20\unit{mg/ml}$ and $c=2\unit{mg/ml}$, respectively).
The insets show the corresponding time evolution of $p_b$.
As can be seen, the FS theoretical predictions, with no fit parameter, become incapable of representing the numerical data when $p_b \gtrsim 0.5$.
This is very clear for the data at $20\unit{mg/ml}$, for which the FS predictions underestimate the distribution of small clusters by more than a factor of two.
Disagreement between theoretical predictions and numerical data at finite times could originate from kinetic pathways and/or due to the presence of closed loops, neglected in the FS approach.
At long times, however, when thermodynamic equilibrium is approached, disagreement can only be ascribed to the presence of intracluster bonds.

To double-check the role of intracluster bonds -- facilitated by the flexibility of the NS and by the growing density of $B$ sites on increasing the cluster size --, we calculate the number of clusters with no reactive $A$ strands, which is equal to the number of intracluster bonds.
We also separate the clusters into two groups: the proper FS loopless clusters and the ones with intracluster bonds.
The latter violate the hypothesis of the FS theory, which assumes that each cluster has one and only one reactive $A$ site.
For the clusters with no loops (NL), we calculate the total number of existing bonds $\NbNL$ and the total number of particles composing these clusters $\NNL$.
The ratio between these two numbers provides the bond probability $\pbFS = \NbNL / \NNL$ for the subset of clusters satisfying the FS hypotheses.
The size distributions of the FS-compatible NL clusters for different simulation times are  shown in Figs.~\ref{fig:csd_20mgml}b and~\ref{fig:csd_2mgml}b and compared with the FS predictions Eq.~\eqref{eq:csd}, identifying $N$ with $\NNL$ and $p_b$ with $\pbFS$.
The quality of the agreement confirms that for the NL clusters, where the FS hypotheses hold by default, the mean-field predictions properly represent the data, suggesting that the time evolution of the aggregation process is sufficiently well-described by the equilibrium solutions~\cite{Corezzi2012}.

\begin{figure}[tb]
\includegraphics[width=0.5\textwidth]{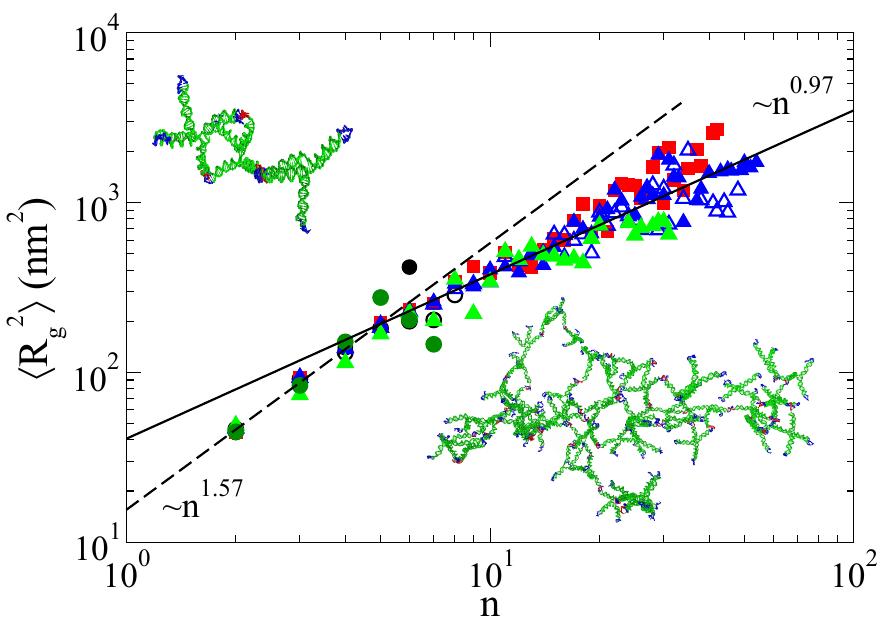}
\caption{Average squared radius of gyration as a function of the cluster size $n$ for different simulations. Dashed lines are power law fits for $n \le 5$ and $n > 5$. Legend: triangles, $c=20\unit{mg/ml}$ for $N=300$ (green) and  $N=2000$ (blue, open and full); red squares, $c=10\unit{mg/ml}$, $N=2000$; circles, $c=2\unit{mg/ml}$ for $N=300$ (dark green) and $N=2000$ (black).  In the top-left portion: a small cluster ($n=3$). In the bottom-right portion: a large cluster ($n=49$).
}
\label{fig:radius_gyration}
\end{figure}

Additionally, we provide a quantification of the structural properties of the clusters.
We evaluate the mean squared radius of gyration
\begin{equation}
\langle R_g^2 \rangle(n) = \frac{1}{2n^2} \left \langle \sum_{j=1}^n \sum_{k=1}^{n} ({\bf r}_j-{\bf r}_k)^2 \right \rangle ,
\end{equation}
where ${\bf r}_j$ is the position of the center of mass of the $j$-th NS belonging to an aggregate of size $n$ and the angular parentheses indicate an ensemble average over all the clusters with the same size $n$ and over time.
For fractal objects,
\begin{equation}
\langle R_g^2 \rangle(n) \sim n^{\frac{2}{d_f}} ,
\end{equation}
where $d_f$ is the fractal exponent.
A power law fit of the data presented in Fig.~\ref{fig:radius_gyration} shows that clusters larger than 3-4 monomers grow with $d_f \approx 2$.

Going back to the cluster size distribution, 
we next provide an extension of the FS theory to account for intracluster bonds. We recall that, formally, for weakly-interacting clusters, the probability of formation of a cluster of size $n$ in equilibrium is proportional~\cite{Hill1987,Sciortino2019} to its partition function $Q_n$ multiplied by a (concentration-dependent) activity $z^n$.
As shown in Section B of the ESI\dag, the FS equation can indeed be recast in this ideal-gas of clusters thermodynamic formalism as
\begin{equation}
\NcNL(n) = \QnNL z^n ,
\end{equation}
where the partition function of a loopless cluster
\begin{equation}
\QnNL = \frac{V}{V_{\text{ref}}} \frac{[(f-1)n]!}{n![(f-2)n+1]!} \left( \E^{-\beta \Fbond} \right)^{n-1}
\end{equation}
is proportional to the system volume, measured in units of a reference volume $V_\text{ref}$, and is composed of a free-energy dependent term $\exp{[-\beta \Fbond (V_{\text{ref}},T)]}$, modelling the contribution of the formation of $n-1$ bonds, and the FS combinatorial entropic term. $\beta = 1/k_{\text{B}} T$ as usual.
The term $z$ plays the role of an activity and its value controls the concentration of the system.
Being $Q_1^{\text{NL}}=V/V_{\text{ref}}$, it is possible to identify $z$ with the nondimensional concentration of \emph{unbonded} particles $\NcNL(1) V_{\text{ref}} / V$.
The mapping between $p_b$ and $\Fbond$ is provided in Section B of the ESI\dag.

To include the possibility of intracluster bonds, we add the partition function associated with the configurations \emph{with} loops as
\begin{equation}
N_c(n) = \left( \QnNL + \QnL \right) z^n .
\label{eq:nnl}
\end{equation}
The partition function $\QnL$, compared to $\QnNL$, must include two terms: (i) an additional factor $\exp{(-\beta \Fbond)}$, which accounts for the presence of the intracluster extra bond, and (ii) a model-dependent factor $g(n,\beta)$, which quantifies the free-energy gain of forming an intracluster bond.
The factor $g(n,\beta)$ includes the relative number of microscopic configurations with an intracluster loop (with respect to a loopless cluster).
It may also include the free-energy cost of bringing the selected $B$ site close to the unbonded $A$ site and thus it could, in principle, (weakly) depend on $T$ if the monomer arms are not quite flexible.
The partition function $\QnL$ can then be written as
\begin{equation}
\QnL = g(n,\beta) \, \QnNL \, \E^{-\beta \Fbond} .
\label{eq:nnl2}
\end{equation}
The unknown $g(n,\beta)$ can be estimated by evaluating the ratio between the number of clusters of size $n$ with and without loops from the simulated configurations at long time, when equilibrium has been reached,
\begin{equation}
\frac{\NcL(n)}{\NcNL(n)} = \frac{\QnL}{\QnNL} = g(n,\beta) \, \E^{-\beta \Fbond} .
\label{eq:rationQm}
\end{equation}
We note on passing that the ratio $\NcL(n) / \NcNL(n)$ depends only on the temperature but not on the concentration.

\begin{figure}[tb]
\includegraphics[width=\columnwidth]{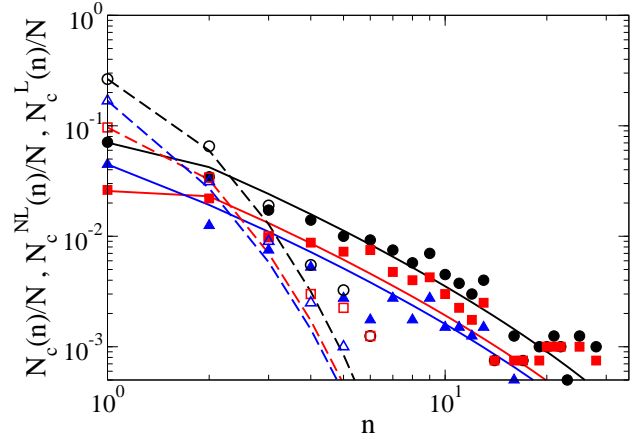}
\caption{Comparison of the prediction of Eqs.~\eqref{eq:nnl} and~\eqref{eq:nnl2} (lines) with the simulation equilibrium data (symbols). The data are obtained by averaging the cluster size distribution from two simulations at the same simulation time. Full symbols refer to the data for $c = 20\unit{mg/ml}$ ($t \simeq 2.85\unit{\mu s} $). 
Open symbols refer to $c = 2\unit{mg/ml}$ ($t \simeq 6.9\unit{\mu s}$). 
Legend: circles, $N_c(n)/N$; squares, $\NcL(n)/N$; triangles, $\NcNL(n)/N$.
}
\label{fig:fit}
\end{figure}

\begin{figure}[tb]
\includegraphics[width=\columnwidth]{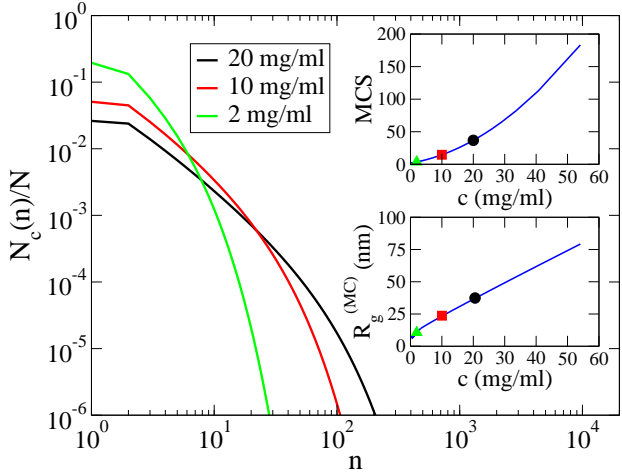}
\caption{Main: Predicted number distribution of clusters $N_c(n)/N$ in the limit of fully bonded system (Eq.~\eqref{eq:lowT}). Top inset: mean cluster size ($\text{MCS} = \sum_n n^2 N_c(n) / \sum_n n N_c(n)$) as a function of the NS concentration. The line is obtained by analytically computing the mean cluster size from the low temperature cluster distribution of Eq.~\eqref{eq:lowT}. Bottom inset: radius of gyration of the mean cluster $R_g^{(\text{MC})}$ as a function of the NS concentration. The line is obtained by combining the MCS dependence with the fit of the radius of gyration shown in Fig.~\ref{fig:radius_gyration}.}
\label{fig:teoriT0}
\end{figure}

Fig.~\ref{fig:fit} compares the predictions of Eqs.~\eqref{eq:nnl} and~\eqref{eq:nnl2} with the numerical data at long times, close to (if not at) equilibrium.
In the comparison, $g(n,\beta)$ and $\exp{(-\beta \Fbond)}$ are the only fit parameters
($\beta$ is fixed), identical for all the densities.
The values of $z$ are fixed by the concentration of loopless monomers.
Despite the intrinsic noise of the data, the theoretical predictions well-represent the numerical values at all densities. The fit suggests that $g(n,\beta)$ is essentially constant already for $n \gtrsim 2$ (\emph{i.e.} for clusters composed by two monomers or more).
This confirms that the unbonded $A$ site essentially binds with a $B$ site on the same particle or, at most, with one of its neighbor monomers.
To provide additional support for this statement, we investigate the distribution of loop sizes\textit{$^{*}$}\footnotetext{\textit{$^{*}$The loop size is defined as the number of monomers in a cluster that form a closed cycle of bonds.}}, confirming that the average loop size is quite small ($\simeq 1.7 \pm 1$).

Even at the coarse-grained level of the oxDNA model, simulations are still too demanding to access lower temperatures than the one we have studied.
However, the previous model allows us to predict the expected cluster size distribution at low-$T$, when the driving force for bonding becomes quite strong and all $A$ sites have reacted.
Under these conditions, $Q^{NL}$ is negligible compared to $Q^L$ for all $n$.
The cluster size distribution will coincide with the distribution of the clusters with an intracluster bond and, therefore, will be given by
\begin{equation}
N_c(n) = \frac{V}{V_{\text{ref}}} g(n,\beta) \frac{[(f-1)n]!}{n![(f-2)n+1]!} \left( \E^{-\beta \Fbond} \right)^n z^n ,
\end{equation}
which can be recast in the form
\begin{equation}
N_c(n) = \frac{V}{V_{\text{ref}}} g(n,\beta) \frac{[(f-1)n]!}{n![(f-2)n+1]!} \left( \frac{V_{\text{ref}}N_c(1)}{V g(1,\beta)} \right)^n .
\label{eq:lowT}
\end{equation}

The term $[V_{\text{ref}}N_c(1)/V g(1,\beta)]$ acts as a renormalized activity. Its value can be tuned to fix the average concentration.
The predicted low-$T$ cluster size distributions for the three different investigated densities are shown in Fig.~\ref{fig:teoriT0}.
The insets of the same figure show the associated mean cluster size (MCS) and the relative radius of gyration of the mean cluster $R_g^{(\text{MC})}$ as a function of the NS concentration, respectively.
From these results, we can formulate three important considerations.
First, Eq.~\eqref{eq:lowT} shows that the temperature (apart from the weak dependence 
entering in $g(n,\beta)$)
 does not play any role: once all possible bonds are formed, the equilibrium distributions are the ones that maximize the entropy.
Second, the same equation shows that the NS concentration modulates the cluster size distribution, at odd with the FS predictions which suggest the formation of an infinite percolating cluster incorporating all monomers.
Third, and more important, the cluster size distribution remains finite at all physical values of the NS concentrations.
Hence, the chance to form intracluster bonds eliminates the possibility to approach the percolation transition.
In a more physical way, Eq.~\eqref{eq:lowT} tells us that, when particles can satisfy all their bonds within the same cluster, the fully bonded (low-$T$) configuration is not the percolating one.
Rather, the equilibrium low-$T$ state exploits the entropic gain provided by the exploration of the system volume by a multiplicity of clusters, modulated by a slightly modified -- by $g(n,\beta)$ -- FS combinatorial term.

\subsection*{Experimental results}
The simulation study has revealed that the presence of intracluster bonds strongly limits the formation of larger clusters in the system, preventing the possibility to reach the percolation point even when all possible bonds are formed ($p_b=1$).
In this case, it has also shown that the cluster size distribution is strongly concentration-dependent, with a MCS (inset of Fig.~\ref{fig:teoriT0}) that is predicted to remain finite at all experimentally accessible NS concentrations.

To test the numerical findings, we realize the very same system in the laboratory and examine it via DLS.
With experiments, we are not limited to the investigation of one single temperature.
Instead, by changing $T$ we can probe different $p_b$ values and even explore the $T$ window where all bonds are formed ($p_b = 1$ for $T \lesssim 20\celsius$), as shown by the melting profile of the sticky sequences reported in Section A of the ESI\dag.
In addition, experiments allow to probe the equilibrium properties of the system.
Samples are let equilibrate for several minutes ($\sim 40\unit{min}$), a time sufficiently long to break and reform several of the bonds between the $A$ and $B$ sticky-ends.
We checked that all results are reproducible upon increasing and decreasing $T$ scans and are not affected by aging nor by the previous history.

\begin{figure*}[tb]
\includegraphics[width=\textwidth,clip]{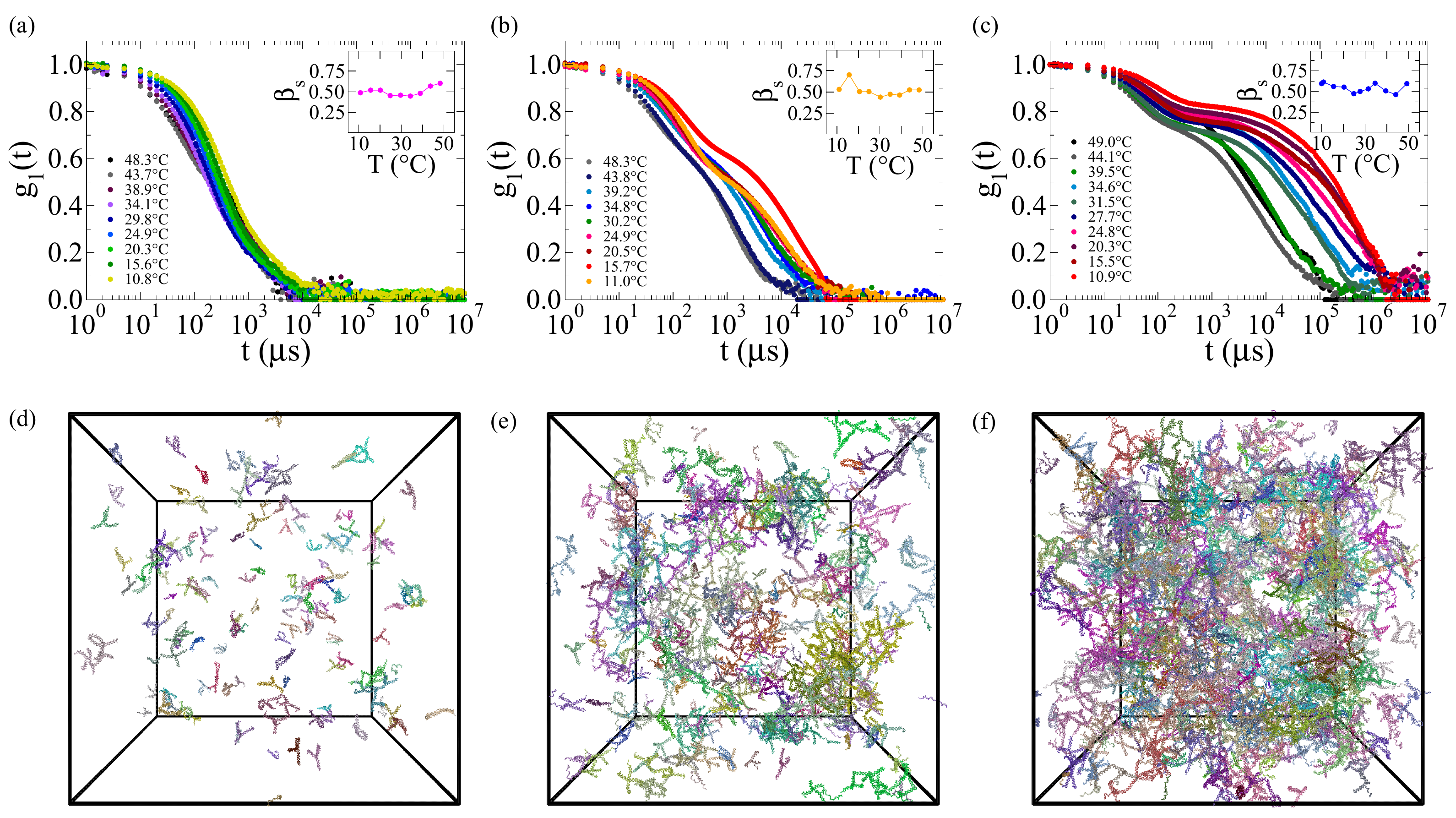}
\caption{Upper panels: DLS results ($g_1(t)$) showing the
$T$-behavior of the system for the three different investigated NS concentrations: (a) $2\unit{mg/ml}$, (b) $10\unit{mg/ml}$, and (c) $20\unit{mg/ml}$. Lower panels: snapshots obtained from simulations performed at the same concentrations of panels a-c. The boxes are displayed on the same scale (box side is $L \simeq 220\unit{nm}$). Different clusters are indicated with different colors.
}
\label{fig:g1_all}
\end{figure*}

Figs.~\ref{fig:g1_all}a-c show the autocorrelation functions of the scattered field $g_1(t)$ for the three investigated NS concentrations ($c = 2\unit{mg/ml}$, $10\unit{mg/ml}$, and $20\unit{mg/ml}$) and for the explored $T$s.
For all samples and temperatures, the correlation functions decay to zero within the experimental accessible time window ($10\unit{s}$), confirming the sample ergodicity. For all studied concentrations, the system if far from a percolation transition, consistent with the
numerical simulations and the proposed theoretical extension of the FS theory.

To quantify the slowing down of the dynamics and to extract a typical (slow) relaxation time, the correlation curves are fitted to a double stretched exponential function (see Section C of the ESI\dag for comparison with the fit function)
\begin{equation}
g_1(t) = (1-A_s) \exp{(-t/\tau_f)} + A_s \exp{(-t/\tau_s)^{\beta_s}} ,
\end{equation}
where $\tau_f$ and $\tau_s$ are the relaxation times of the fast and slow relaxation processes, respectively, $A_s$ is the amplitude of the slow process, and $\beta_s$ its stretching exponent.
The slow relaxation time is better represented by its average value, defined as 
\begin{equation}
\langle\tau_s\rangle=\frac{\int_0^{\infty}t \exp{(-t/\tau_s)^{\beta_s}} \, \textrm{d}t}{\int_0^{\infty}\exp{(-t/\tau_s)^{\beta_s}} \, \textrm{d}t} =\frac{\tau_s}{\beta_s} \,\Gamma\left(\frac{1}{\beta_s}\right) ,   
\end{equation}
where $\Gamma$ is the gamma function.

\begin{figure}[h]
\includegraphics[width=\columnwidth,clip]{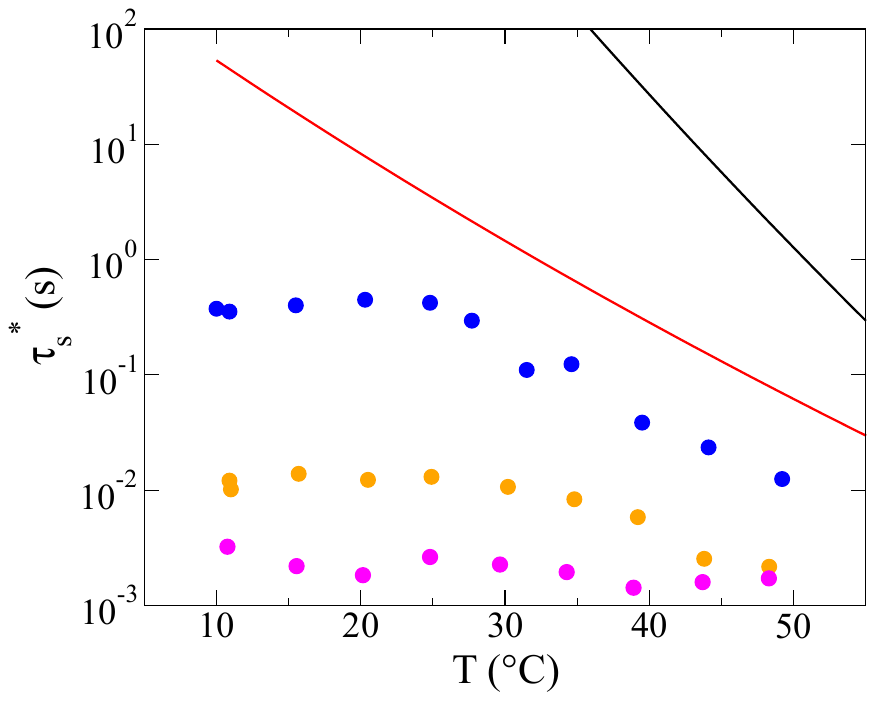}
\caption{$T$-dependence of the (viscosity rescaled) slow decay time $\tau_s^*$ for samples at the three different concentrations: $20\unit{mg/ml}$ (blue), $10\unit{mg/ml}$ (orange), and $2\unit{mg/ml}$ (magenta). The lines are the two estimated bond lifetimes, calculated according to Eq.~\eqref{eq:time} using $\alpha = 0.5$ (red) and $\alpha = 1$ (black), 
to provide support to the assumption that the relaxation process takes place at fixed 
bonding pattern.
}
\label{fig:time}
\end{figure}

The insets of Figs.~\ref{fig:g1_all}a-c display the values of $\beta_s$, which is associated with the slow relaxation process.
For all the measurements, the values lie within the range $0.4 \lesssim \beta_s \lesssim 0.6$.
The values of the slow relaxation time are reported in Fig.~\ref{fig:time} for all the concentrations and temperatures.
To eliminate the trivial effect of the temperature dependence of the solvent viscosity, the times are rescaled to the viscosity $\eta_{\text{solv}}$ of the NaCl $250\unit{mM}$ solvent at the highest investigated temperature ($T_{\text{ref}} \simeq 48.5\celsius$) as
\begin{equation}
  \tau_s^* (T) = \langle\tau_s\rangle (T) \frac{\eta_{\text{solv}}(T_{\text{ref}})}{\eta_{\text{solv}} (T)} .
\end{equation}

In general, two distinct processes -- which can only be disentangled in particular conditions -- contribute to the  relaxation process probed by DLS: cluster \emph{restructuring} and cluster \emph{diffusion}.
The cluster restructuring times, which are related to the breaking and coalescence of different aggregates, are slaved to the bond-breaking times.
In our DNA design, explicitly selected to generate a fully bonded system for $T \lesssim 20\celsius$, 9 base pairs take part to the $AB$ bond, resulting in an enthalpic $\Delta H$ and entropic $\Delta S$ contribution to the binding free-energy $\Delta G$ equal to $\Delta H = -61.5\unit{kcal/mol}$ and $\Delta S = -178.3\unit{cal/mol \, K}$, evaluated using the web application DINAMelt~\cite{Markham2005} based on SantaLucia~\cite{SantaLucia1996}.
A reasonable estimate of the bond lifetime is~\cite{Bomboi2015}
\begin{equation}
\tau = \tau_0 \, \E^{\alpha \Delta G/RT} ,
\label{eq:time}
\end{equation}
where $\tau_0$ is of the order of a few ms and $R = 1.987\unit{cal/mol \, K}$.
The value of the coefficient $\alpha$ is between 0.5 and 2, depending on the degree of cooperativity of the bond-breaking process~\cite{Smallenburg2013}.
Here, we use $\tau_0 = 3\unit{ms}$, as found in Ref.~\cite{Bomboi2015}. In Fig.~\ref{fig:time}, we show the expected $T$-dependence of the bond lifetime as predicted by Eq.~\eqref{eq:time} for $\alpha=0.5$ and $\alpha=1$.
In both cases, the experimental times are faster than the bond-breaking time, even when the most conservative estimate of the bond lifetime is selected ($\alpha=0.5$).
This leaves the cluster diffusion as the main mechanism for the decorrelation of the density fluctuations in this system.

At the lowest density ($c = 2\unit{mg/ml}$, Fig.~\ref{fig:g1_all}a), where the hypothesis of independent clusters is more realistic, the relaxation process is quite fast and it does not show a clear hint of two-step relaxation, suggesting that the experiments are probing the free diffusion of limited-size clusters.
The relaxation time grows only by a factor of $\approx 5$ when $T$ varies from $50\celsius$ to $10\celsius$ (corresponding to $p_b$ from $\approx 0$ to $\approx 1$).
At $c = 10\unit{mg/ml}$ (Fig.~\ref{fig:g1_all}b), a weak additional relaxation process appears, signalling the onset of interactions between clusters.
The image from the simulation (Fig.~\ref{fig:g1_all}e), which we can use as a visual aid to interpret the DLS data, suggests that it may become appropriate to tentatively interpret the fast decay as originating from the clusters free diffusion, while the slow one comes from the confinement induced by the presence of nearby clusters.
This hypothesis is supported by the data at $c = 20\unit{mg/ml}$ (Fig.~\ref{fig:g1_all}c) -- and again  by the simulation snapshot in Fig.~\ref{fig:g1_all}f --, where the excluded-volume interactions between clusters are significant, as illustrated by the pronounced two-step decay of the correlation functions. 

We then intend to compare the simulation results with the experiments for sample at $c = 2\unit{mg/ml}$.
We evaluate the intensity-averaged$^*$\footnotetext{$^*$~The intensity average for the simulation data is obtained by performing the average weighted by distribution $n^2 N_c(n)$.} hydrodynamic radius from the cluster size distribution obtained from the simulations at $T = 45\celsius$ (\emph{cf.} Fig.~\ref{fig:fit}) and from the low-$T$ theory (\emph{cf.} Fig.~\ref{fig:teoriT0}).
For each cluster of size $n$, we calculate the hydrodynamic radius following the recipe described in Ref.~\cite{Rovigatti2019}, which provides the equivalent ellipsoid associated with the cluster.
The actual hydrodynamic radius is then computed from known formulae~\cite{Hubbard1993}. 
Finally, as done for the radius of gyration, we evaluate the cluster size dependence of the ensemble average of $R_h$.
The power law fit to the data gives $\langle R_h \rangle (n) \simeq R_0 \, n^{\gamma}$, with 
\begin{equation}
R_0 \simeq
\begin{cases}
5.0 \unit{nm} \quad \text{if } n\le 5 \\
6.5 \unit{nm} \quad \text{if } n > 5 
\end{cases}
\end{equation}
and
\begin{equation}
\gamma =
\begin{cases}
0.68 \quad \text{if } n\le 5 \\
0.53 \quad \text{if } n > 5.
\end{cases}
\end{equation}

For further details, see Section D of the ESI\dag.
At high temperature, equal to the one numerically investigated ($T = 45\celsius$), we obtain from the simulations an intensity-averaged hydrodynamic radius of $\langle R_h \rangle_z \simeq 9.3\unit{nm}$ (using the radius of gyration, one obtains $\langle R_g \rangle_z \simeq 8.1\unit{nm}$).
By analyzing the correlation time associated with the fast relaxation $\tau_f$, we estimate a mean hydrodynamic radius $R_h \simeq 5.5\unit{nm}$. 
At low temperature, instead, the numerical prediction (see Fig.~\ref{fig:teoriT0}) provides an expected radius $\langle R_h \rangle_z \simeq 14.9\unit{nm}$ ($\langle R_g \rangle_z \simeq 13.2\unit{nm}$), to be compared to the experimental value of $R_h \simeq 22.5\unit{nm}$. 

The discrepancy between the experimental values and the ones estimated from the simulations -- which is about 50\% at low temperature and grows to a factor of  $\sim 0.7$ at $T=45\celsius$ --, can be mainly ascribed to the approximate model we used for the calculation of the hydrodynamic radius. Another possible source of error can be related to the numerical noise of the data used in the fit to obtain $g(n,\beta)$. 


\section*{Conclusions}

This article, based on a combined numerical, theoretical, and experimental study, reports numerous relevant findings.

\begin{itemize}
\item
It shows that DNA-made NSs with precise binding topologies can be nowadays produced in bulk quantities and exploited as highly precise model systems to tackle the physics of aggregating particles, either directional colloids or functional polymers. 
Here, we studied the aggregation behavior of $AB_3$ monomers made of DNA NSs with the aim of clarifying the behavior of hyperbranched aggregation when intracluster bonds are present.

\item
It reports an extensive MD simulations  study, based on a high-quality coarse-grained potential for the DNA interactions, to investigate the equilibrium cluster size distribution at fixed temperature and at different initial monomer concentrations.
We found that the mean cluster size increases with the particle concentration, from a suspension of monomers and small clusters, at low $c$, to a highly polydisperse solution at large concentrations.
Simulations also suggested that the presence of intracluster bonds eliminates the possibility to reach the percolation transition.

\item 
It reports a novel theoretical way to include the presence of intracluster bonds in the formalism. First, the FS approach is translated in an ideal-gas of clusters formalism,
 which is then extended to include also clusters with intracluster bonds.
 Including a model-dependent  (but concentration-independent)
quantity $g(n,\beta)$ -- which can be determined 
 as a fit parameter --, it becomes possible to predict
 the cluster size distribution without limiting assumptions. 
 The theory explains why percolation is avoided when 
 intracluster bonds are possible. 
 Comparison with MD simulation data supports the
 quality of the modelling. 

\item It reports  DLS measurements of the same system studied numerically and theoretically
to provide evidence that, at odd with the FS predicitions,  percolation is not encountered in this
system.   Indeed, by decreasing the temperature, and exploiting the reversibility of the sticky-end pairing, 
it is possible to investigate the effect of  bonding, up to the point where all bonds in the system are formed.
The data we collected strongly support the idea that the polydispersity of the aggregates remains finite at low temperatures.
Additionally, it suggests that the equilibrium cluster size distribution only depends on the initial monomer concentration.
\end{itemize}
We believe that these results will also be valuable for the community interested in the biotechnological applications of hyperbranched polymers, due to the biocompatibility and versatility of DNA as a building-block for complex and innovative materials.

\section*{Conflicts of interest}
There are no conflicts to declare.

\section*{Acknowledgements}
We acknowledge support  
from  MIUR-PRIN  (Grant No. 2017Z55KCW) and
from Regione Lazio (Grant No. 85857-0051-0085).
We thank L. Rovigatti for his help with the oxDNA code.



\balance


\bibliography{bibliography} 

\providecommand*{\mcitethebibliography}{\thebibliography}
\csname @ifundefined\endcsname{endmcitethebibliography}
{\let\endmcitethebibliography\endthebibliography}{}
\begin{mcitethebibliography}{54}
\providecommand*{\natexlab}[1]{#1}
\providecommand*{\mciteSetBstSublistMode}[1]{}
\providecommand*{\mciteSetBstMaxWidthForm}[2]{}
\providecommand*{\mciteBstWouldAddEndPuncttrue}
  {\def\EndOfBibitem{\unskip.}}
\providecommand*{\mciteBstWouldAddEndPunctfalse}
  {\let\EndOfBibitem\relax}
\providecommand*{\mciteSetBstMidEndSepPunct}[3]{}
\providecommand*{\mciteSetBstSublistLabelBeginEnd}[3]{}
\providecommand*{\EndOfBibitem}{}
\mciteSetBstSublistMode{f}
\mciteSetBstMaxWidthForm{subitem}
{(\emph{\alph{mcitesubitemcount}})}
\mciteSetBstSublistLabelBeginEnd{\mcitemaxwidthsubitemform\space}
{\relax}{\relax}

\bibitem[Bianchi \emph{et~al.}(2007)Bianchi, Tartaglia, Nave, and
  Sciortino]{Bianchi2007}
E.~Bianchi, P.~Tartaglia, E.~L. Nave and F.~Sciortino, \emph{J. Chem. Phys. B},
  2007, \textbf{111}, 11765\relax
\mciteBstWouldAddEndPuncttrue
\mciteSetBstMidEndSepPunct{\mcitedefaultmidpunct}
{\mcitedefaultendpunct}{\mcitedefaultseppunct}\relax
\EndOfBibitem
\bibitem[Corezzi \emph{et~al.}(2008)Corezzi, Michele, Zaccarelli, Fioretto, and
  Sciortino]{Corezzi2008}
S.~Corezzi, C.~D. Michele, E.~Zaccarelli, D.~Fioretto and F.~Sciortino,
  \emph{Soft Matter}, 2008, \textbf{4}, 1173\relax
\mciteBstWouldAddEndPuncttrue
\mciteSetBstMidEndSepPunct{\mcitedefaultmidpunct}
{\mcitedefaultendpunct}{\mcitedefaultseppunct}\relax
\EndOfBibitem
\bibitem[Wertheim(1984)]{Wertheim1984}
M.~Wertheim, \emph{J. Stat. Phys.}, 1984, \textbf{35}, 19--34\relax
\mciteBstWouldAddEndPuncttrue
\mciteSetBstMidEndSepPunct{\mcitedefaultmidpunct}
{\mcitedefaultendpunct}{\mcitedefaultseppunct}\relax
\EndOfBibitem
\bibitem[Wertheim(1986)]{Wertheim1986}
M.~Wertheim, \emph{J. Stat. Phys.}, 1986, \textbf{42}, 459--476\relax
\mciteBstWouldAddEndPuncttrue
\mciteSetBstMidEndSepPunct{\mcitedefaultmidpunct}
{\mcitedefaultendpunct}{\mcitedefaultseppunct}\relax
\EndOfBibitem
\bibitem[Flory(1941)]{Flory1941}
P.~Flory, \emph{J. Am. Chem. Soc.}, 1941, \textbf{63}, 3083\relax
\mciteBstWouldAddEndPuncttrue
\mciteSetBstMidEndSepPunct{\mcitedefaultmidpunct}
{\mcitedefaultendpunct}{\mcitedefaultseppunct}\relax
\EndOfBibitem
\bibitem[Stockmayer(1943)]{Stockmayer1943}
W.~Stockmayer, \emph{J. Chem. Phys.}, 1943, \textbf{11}, 45--55\relax
\mciteBstWouldAddEndPuncttrue
\mciteSetBstMidEndSepPunct{\mcitedefaultmidpunct}
{\mcitedefaultendpunct}{\mcitedefaultseppunct}\relax
\EndOfBibitem
\bibitem[Bianchi \emph{et~al.}(2006)Bianchi, Largo, Tartaglia, Zaccarelli, and
  Sciortino]{Bianchi2006}
E.~Bianchi, J.~Largo, P.~Tartaglia, E.~Zaccarelli and F.~Sciortino, \emph{Phys.
  Rev. Lett.}, 2006, \textbf{97}, 168301\relax
\mciteBstWouldAddEndPuncttrue
\mciteSetBstMidEndSepPunct{\mcitedefaultmidpunct}
{\mcitedefaultendpunct}{\mcitedefaultseppunct}\relax
\EndOfBibitem
\bibitem[Sciortino \emph{et~al.}(2007)Sciortino, Bianchi, Douglas, and
  Tartaglia]{Sciortino2007}
F.~Sciortino, E.~Bianchi, J.~Douglas and P.~Tartaglia, \emph{J. Chem. Phys.},
  2007, \textbf{126}, 194903\relax
\mciteBstWouldAddEndPuncttrue
\mciteSetBstMidEndSepPunct{\mcitedefaultmidpunct}
{\mcitedefaultendpunct}{\mcitedefaultseppunct}\relax
\EndOfBibitem
\bibitem[Kim and Webster(1990)]{Kim1990}
Y.~Kim and O.~Webster, \emph{J. Am. Chem. Soc.}, 1990, \textbf{112}, 4592\relax
\mciteBstWouldAddEndPuncttrue
\mciteSetBstMidEndSepPunct{\mcitedefaultmidpunct}
{\mcitedefaultendpunct}{\mcitedefaultseppunct}\relax
\EndOfBibitem
\bibitem[Kim and Webster(1992)]{Kim1992}
Y.~Kim and O.~Webster, \emph{Macromolecules}, 1992, \textbf{25}, 5561\relax
\mciteBstWouldAddEndPuncttrue
\mciteSetBstMidEndSepPunct{\mcitedefaultmidpunct}
{\mcitedefaultendpunct}{\mcitedefaultseppunct}\relax
\EndOfBibitem
\bibitem[Cuneo and Gao(2020)]{Cuneo2020}
T.~Cuneo and H.~Gao, \emph{WIREs Nanomed Nanobiotechnol.}, 2020,  e1640\relax
\mciteBstWouldAddEndPuncttrue
\mciteSetBstMidEndSepPunct{\mcitedefaultmidpunct}
{\mcitedefaultendpunct}{\mcitedefaultseppunct}\relax
\EndOfBibitem
\bibitem[Jochum \emph{et~al.}(2019)Jochum, Ad{\v{z}}i{\'c}, Stiakakis, Derrien,
  Luo, Kahl, and Likos]{Jochum2019}
C.~Jochum, N.~Ad{\v{z}}i{\'c}, E.~Stiakakis, T.~Derrien, D.~Luo, G.~Kahl and
  C.~Likos, \emph{Nanoscale}, 2019, \textbf{11}, 1604--1617\relax
\mciteBstWouldAddEndPuncttrue
\mciteSetBstMidEndSepPunct{\mcitedefaultmidpunct}
{\mcitedefaultendpunct}{\mcitedefaultseppunct}\relax
\EndOfBibitem
\bibitem[Zhou \emph{et~al.}(2010)Zhou, Huang, Liu, Zhu, and Yan]{Zhou2010}
B.~Zhou, W.~Huang, J.~Liu, X.~Zhu and D.~Yan, \emph{Adv. Mater.}, 2010,
  \textbf{22}, 4567\relax
\mciteBstWouldAddEndPuncttrue
\mciteSetBstMidEndSepPunct{\mcitedefaultmidpunct}
{\mcitedefaultendpunct}{\mcitedefaultseppunct}\relax
\EndOfBibitem
\bibitem[Liu \emph{et~al.}(2015)Liu, Huang, Pang, and Yan]{Liu2015}
J.~Liu, W.~Huang, Y.~Pang and D.~Yan, \emph{Chem. Soc. Rev.}, 2015,
  \textbf{44}, 3942\relax
\mciteBstWouldAddEndPuncttrue
\mciteSetBstMidEndSepPunct{\mcitedefaultmidpunct}
{\mcitedefaultendpunct}{\mcitedefaultseppunct}\relax
\EndOfBibitem
\bibitem[van Benthem(2000)]{vanBenthem2000}
R.~van Benthem, \emph{Prog. Org. Coat.}, 2000, \textbf{40}, 203--214\relax
\mciteBstWouldAddEndPuncttrue
\mciteSetBstMidEndSepPunct{\mcitedefaultmidpunct}
{\mcitedefaultendpunct}{\mcitedefaultseppunct}\relax
\EndOfBibitem
\bibitem[Zotti \emph{et~al.}(2019)Zotti, Zuppolini, Borriello, and
  Zarrelli]{Zotti2019}
A.~Zotti, S.~Zuppolini, A.~Borriello and M.~Zarrelli, \emph{Nanomaterials},
  2019, \textbf{9}, 418\relax
\mciteBstWouldAddEndPuncttrue
\mciteSetBstMidEndSepPunct{\mcitedefaultmidpunct}
{\mcitedefaultendpunct}{\mcitedefaultseppunct}\relax
\EndOfBibitem
\bibitem[Paleos \emph{et~al.}(2010)Paleos, Tsiourvas, Sideratou, and
  Tziveleka]{Paleos2010}
C.~Paleos, D.~Tsiourvas, Z.~Sideratou and L.-A. Tziveleka, \emph{Expert Opin.
  Drug Deliv.}, 2010, \textbf{7}, 1387\relax
\mciteBstWouldAddEndPuncttrue
\mciteSetBstMidEndSepPunct{\mcitedefaultmidpunct}
{\mcitedefaultendpunct}{\mcitedefaultseppunct}\relax
\EndOfBibitem
\bibitem[Gajbhiye \emph{et~al.}(2013)Gajbhiye, Escalante, Chen, Laperle, Zheng,
  Steyer, Gong, and Saha]{Gajbhiye2013}
V.~Gajbhiye, L.~Escalante, G.~Chen, A.~Laperle, Q.~Zheng, B.~Steyer, S.~Gong
  and K.~Saha, \emph{Nanoscale}, 2013, \textbf{6}, 521--531\relax
\mciteBstWouldAddEndPuncttrue
\mciteSetBstMidEndSepPunct{\mcitedefaultmidpunct}
{\mcitedefaultendpunct}{\mcitedefaultseppunct}\relax
\EndOfBibitem
\bibitem[Qi \emph{et~al.}(2016)Qi, Duan, Yu, Yao, Tian, and Xu]{Qi2016}
M.~Qi, S.~Duan, B.~Yu, H.~Yao, W.~Tian and F.~Xu, \emph{Polym. Chem.}, 2016,
  \textbf{7}, 4334\relax
\mciteBstWouldAddEndPuncttrue
\mciteSetBstMidEndSepPunct{\mcitedefaultmidpunct}
{\mcitedefaultendpunct}{\mcitedefaultseppunct}\relax
\EndOfBibitem
\bibitem[Zhao \emph{et~al.}(2017)Zhao, Wu, Wang, Duan, Wang, Punjabi, Zhao,
  Zhang, Xu, Gao, and Han]{Zhao2017}
L.~Zhao, X.~Wu, X.~Wang, C.~Duan, H.~Wang, A.~Punjabi, Y.~Zhao, Y.~Zhang,
  Z.~Xu, H.~Gao and G.~Han, \emph{ACS Macro Lett.}, 2017, \textbf{6},
  700--704\relax
\mciteBstWouldAddEndPuncttrue
\mciteSetBstMidEndSepPunct{\mcitedefaultmidpunct}
{\mcitedefaultendpunct}{\mcitedefaultseppunct}\relax
\EndOfBibitem
\bibitem[Qi \emph{et~al.}(2019)Qi, Duan, Yu, Yao, Tian, and Xu]{Song2019}
M.~Qi, S.~Duan, B.~Yu, H.~Yao, W.~Tian and F.~Xu, \emph{Colloids Surf. B},
  2019, \textbf{182}, 110375\relax
\mciteBstWouldAddEndPuncttrue
\mciteSetBstMidEndSepPunct{\mcitedefaultmidpunct}
{\mcitedefaultendpunct}{\mcitedefaultseppunct}\relax
\EndOfBibitem
\bibitem[Feng \emph{et~al.}(2012)Feng, Ding, and Liu]{Feng2012}
G.~Feng, D.~Ding and B.~Liu, \emph{Nanoscale}, 2012, \textbf{4}, 6150\relax
\mciteBstWouldAddEndPuncttrue
\mciteSetBstMidEndSepPunct{\mcitedefaultmidpunct}
{\mcitedefaultendpunct}{\mcitedefaultseppunct}\relax
\EndOfBibitem
\bibitem[Malekzadeh \emph{et~al.}(2017)Malekzadeh, Ramazani, Rezaei, and
  Niknejad]{MashhadiMalekzadeh2017}
A.~M. Malekzadeh, A.~Ramazani, S.~T. Rezaei and H.~Niknejad, \emph{J. Colloid
  Interface Sci.}, 2017, \textbf{490}, 64--73\relax
\mciteBstWouldAddEndPuncttrue
\mciteSetBstMidEndSepPunct{\mcitedefaultmidpunct}
{\mcitedefaultendpunct}{\mcitedefaultseppunct}\relax
\EndOfBibitem
\bibitem[Pitois \emph{et~al.}(2001)Pitois, Wiesmann, Lindgren, and
  Hult]{Pitois2001}
C.~Pitois, D.~Wiesmann, M.~Lindgren and A.~Hult, \emph{Adv. Mater.}, 2001,
  \textbf{13}, 1483\relax
\mciteBstWouldAddEndPuncttrue
\mciteSetBstMidEndSepPunct{\mcitedefaultmidpunct}
{\mcitedefaultendpunct}{\mcitedefaultseppunct}\relax
\EndOfBibitem
\bibitem[Shi \emph{et~al.}(2013)Shi, Chen, Liu, Xu, An, Ouyang, Tu, Zhao, Fan,
  Wang, and Huang]{Shi2013}
H.~Shi, X.~Chen, S.~Liu, H.~Xu, Z.~An, L.~Ouyang, Z.~Tu, Q.~Zhao, Q.~Fan,
  L.~Wang and W.~Huang, \emph{ACS Appl. Mater. Interfaces}, 2013, \textbf{5},
  4562\relax
\mciteBstWouldAddEndPuncttrue
\mciteSetBstMidEndSepPunct{\mcitedefaultmidpunct}
{\mcitedefaultendpunct}{\mcitedefaultseppunct}\relax
\EndOfBibitem
\bibitem[Liu \emph{et~al.}(2014)Liu, Liu, Cheng, and Chen]{Liu2014}
X.~Liu, H.-J. Liu, F.~Cheng and Y.~Chen, \emph{Nanoscale}, 2014, \textbf{6},
  7453\relax
\mciteBstWouldAddEndPuncttrue
\mciteSetBstMidEndSepPunct{\mcitedefaultmidpunct}
{\mcitedefaultendpunct}{\mcitedefaultseppunct}\relax
\EndOfBibitem
\bibitem[Zheng \emph{et~al.}(2011)Zheng, Cao, Newland, Dong, Pandit, and
  Wang]{Zheng2011}
Y.~Zheng, H.~Cao, B.~Newland, Y.~Dong, A.~Pandit and W.~Wang, \emph{J. Am.
  Chem. Soc.}, 2011, \textbf{133}, 13130\relax
\mciteBstWouldAddEndPuncttrue
\mciteSetBstMidEndSepPunct{\mcitedefaultmidpunct}
{\mcitedefaultendpunct}{\mcitedefaultseppunct}\relax
\EndOfBibitem
\bibitem[Lyu \emph{et~al.}(2018)Lyu, Gao, Zhang, Greiser, Tai, and
  Wang]{Lyu2018}
J.~Lyu, Y.~Gao, Z.~Zhang, U.~Greiser, H.~Tai and W.~Wang, \emph{Sci. China
  Chem.}, 2018, \textbf{61}, 319--327\relax
\mciteBstWouldAddEndPuncttrue
\mciteSetBstMidEndSepPunct{\mcitedefaultmidpunct}
{\mcitedefaultendpunct}{\mcitedefaultseppunct}\relax
\EndOfBibitem
\bibitem[Biffi \emph{et~al.}(2013)Biffi, Cerbino, Bomboi, Paraboschi, Asselta,
  Sciortino, and Bellini]{Biffi2013}
S.~Biffi, R.~Cerbino, F.~Bomboi, E.~Paraboschi, R.~Asselta, F.~Sciortino and
  T.~Bellini, \emph{Proc. Natl. Acad. Sci. U. S. A.}, 2013, \textbf{110},
  15633\relax
\mciteBstWouldAddEndPuncttrue
\mciteSetBstMidEndSepPunct{\mcitedefaultmidpunct}
{\mcitedefaultendpunct}{\mcitedefaultseppunct}\relax
\EndOfBibitem
\bibitem[Biffi \emph{et~al.}(2015)Biffi, Cerbino, Nava, Bomboi, Sciortino, and
  Bellini]{Biffi2015}
S.~Biffi, R.~Cerbino, G.~Nava, F.~Bomboi, F.~Sciortino and T.~Bellini,
  \emph{Soft Matter}, 2015, \textbf{11}, 3132\relax
\mciteBstWouldAddEndPuncttrue
\mciteSetBstMidEndSepPunct{\mcitedefaultmidpunct}
{\mcitedefaultendpunct}{\mcitedefaultseppunct}\relax
\EndOfBibitem
\bibitem[Rovigatti \emph{et~al.}(2014)Rovigatti, Bomboi, and
  Sciortino]{Rovigatti2014}
L.~Rovigatti, F.~Bomboi and F.~Sciortino, \emph{J. Chem. Phys.}, 2014,
  \textbf{140}, 154903\relax
\mciteBstWouldAddEndPuncttrue
\mciteSetBstMidEndSepPunct{\mcitedefaultmidpunct}
{\mcitedefaultendpunct}{\mcitedefaultseppunct}\relax
\EndOfBibitem
\bibitem[Conrad \emph{et~al.}(2019)Conrad, Kennedy, Fygenson, and
  Saleh]{Conrad2019}
N.~Conrad, T.~Kennedy, D.~Fygenson and O.~Saleh, \emph{Proc. Natl. Acad. Sci.
  U. S. A.}, 2019, \textbf{116}, 7238\relax
\mciteBstWouldAddEndPuncttrue
\mciteSetBstMidEndSepPunct{\mcitedefaultmidpunct}
{\mcitedefaultendpunct}{\mcitedefaultseppunct}\relax
\EndOfBibitem
\bibitem[Ouldridge \emph{et~al.}(2011)Ouldridge, Louis, and
  Doye]{Ouldridge2011}
T.~Ouldridge, A.~Louis and J.~Doye, \emph{J. Chem. Phys.}, 2011, \textbf{134},
  085101\relax
\mciteBstWouldAddEndPuncttrue
\mciteSetBstMidEndSepPunct{\mcitedefaultmidpunct}
{\mcitedefaultendpunct}{\mcitedefaultseppunct}\relax
\EndOfBibitem
\bibitem[Snodin \emph{et~al.}(2015)Snodin, Randisi, Mosayebi, \v{S}ulc,
  Schreck, Romano, Ouldridge, Tsukanov, Nir, Louis, and Doye]{Snodin2015}
B.~Snodin, F.~Randisi, M.~Mosayebi, P.~\v{S}ulc, J.~Schreck, F.~Romano,
  T.~Ouldridge, R.~Tsukanov, E.~Nir, A.~Louis and J.~Doye, \emph{J. Chem.
  Phys.}, 2015, \textbf{142}, 234901\relax
\mciteBstWouldAddEndPuncttrue
\mciteSetBstMidEndSepPunct{\mcitedefaultmidpunct}
{\mcitedefaultendpunct}{\mcitedefaultseppunct}\relax
\EndOfBibitem
\bibitem[Seeman(2003)]{Seeman2003}
N.~Seeman, \emph{Nature}, 2003, \textbf{421}, 427--431\relax
\mciteBstWouldAddEndPuncttrue
\mciteSetBstMidEndSepPunct{\mcitedefaultmidpunct}
{\mcitedefaultendpunct}{\mcitedefaultseppunct}\relax
\EndOfBibitem
\bibitem[Seeman(2005)]{Seeman2005}
N.~Seeman, in \emph{NanoBiotechnology Protocols}, Springer, 2005, pp.
  143--166\relax
\mciteBstWouldAddEndPuncttrue
\mciteSetBstMidEndSepPunct{\mcitedefaultmidpunct}
{\mcitedefaultendpunct}{\mcitedefaultseppunct}\relax
\EndOfBibitem
\bibitem[Bellini \emph{et~al.}(2011)Bellini, Cerbino, and
  Zanchetta]{Bellini2011}
T.~Bellini, R.~Cerbino and G.~Zanchetta, in \emph{Liquid Crystals}, Springer,
  2011, pp. 225--279\relax
\mciteBstWouldAddEndPuncttrue
\mciteSetBstMidEndSepPunct{\mcitedefaultmidpunct}
{\mcitedefaultendpunct}{\mcitedefaultseppunct}\relax
\EndOfBibitem
\bibitem[Salamonczyk \emph{et~al.}(2016)Salamonczyk, Zhang, Portale, Zhu,
  Kentzinger, Gleeson, Jakli, Michele, Dhont, Sprunt, and
  Stiakakis]{Salamonczyk2016}
M.~Salamonczyk, J.~Zhang, G.~Portale, C.~Zhu, E.~Kentzinger, J.~Gleeson,
  A.~Jakli, C.~D. Michele, J.~Dhont, S.~Sprunt and E.~Stiakakis, \emph{Nat.
  Commun.}, 2016, \textbf{7}, 13358\relax
\mciteBstWouldAddEndPuncttrue
\mciteSetBstMidEndSepPunct{\mcitedefaultmidpunct}
{\mcitedefaultendpunct}{\mcitedefaultseppunct}\relax
\EndOfBibitem
\bibitem[Fernandez-Castanon \emph{et~al.}(2016)Fernandez-Castanon, Bomboi,
  Rovigatti, Zanatta, Paciaroni, Comez, Porcar, Jafta, Fadda, Bellini, and
  Sciortino]{FernandezCastanon2016}
J.~Fernandez-Castanon, F.~Bomboi, L.~Rovigatti, M.~Zanatta, A.~Paciaroni,
  L.~Comez, L.~Porcar, C.~Jafta, G.~Fadda, T.~Bellini and F.~Sciortino,
  \emph{J. Chem. Phys.}, 2016, \textbf{145}, 084910\relax
\mciteBstWouldAddEndPuncttrue
\mciteSetBstMidEndSepPunct{\mcitedefaultmidpunct}
{\mcitedefaultendpunct}{\mcitedefaultseppunct}\relax
\EndOfBibitem
\bibitem[Nguyen and Saleh(2017)]{Nguyen2017}
D.~Nguyen and O.~Saleh, \emph{Soft Matter}, 2017, \textbf{13}, 5421\relax
\mciteBstWouldAddEndPuncttrue
\mciteSetBstMidEndSepPunct{\mcitedefaultmidpunct}
{\mcitedefaultendpunct}{\mcitedefaultseppunct}\relax
\EndOfBibitem
\bibitem[Bomboi \emph{et~al.}(2015)Bomboi, Biffi, Cerbino, Bellini, Bordi, and
  Sciortino]{Bomboi2015}
F.~Bomboi, S.~Biffi, R.~Cerbino, T.~Bellini, F.~Bordi and F.~Sciortino,
  \emph{Eur. Phys. J. E}, 2015, \textbf{38}, 64\relax
\mciteBstWouldAddEndPuncttrue
\mciteSetBstMidEndSepPunct{\mcitedefaultmidpunct}
{\mcitedefaultendpunct}{\mcitedefaultseppunct}\relax
\EndOfBibitem
\bibitem[SantaLucia(1998)]{SantaLucia1998}
J.~SantaLucia, \emph{Proc. Natl. Acad. Sci. U.S.A.}, 1998, \textbf{95},
  1460\relax
\mciteBstWouldAddEndPuncttrue
\mciteSetBstMidEndSepPunct{\mcitedefaultmidpunct}
{\mcitedefaultendpunct}{\mcitedefaultseppunct}\relax
\EndOfBibitem
\bibitem[Holbrook \emph{et~al.}(1999)Holbrook, Capp, Saecker, and
  Record]{Holbrook1999}
J.~A. Holbrook, M.~W. Capp, R.~M. Saecker and M.~T. Record,
  \emph{Biochemistry}, 1999, \textbf{38}, 8409\relax
\mciteBstWouldAddEndPuncttrue
\mciteSetBstMidEndSepPunct{\mcitedefaultmidpunct}
{\mcitedefaultendpunct}{\mcitedefaultseppunct}\relax
\EndOfBibitem
\bibitem[oxd()]{oxdna-code}
\url{https://dna.physics.ox.ac.uk}\relax
\mciteBstWouldAddEndPuncttrue
\mciteSetBstMidEndSepPunct{\mcitedefaultmidpunct}
{\mcitedefaultendpunct}{\mcitedefaultseppunct}\relax
\EndOfBibitem
\bibitem[Russo \emph{et~al.}(2009)Russo, Tartaglia, and Sciortino]{Russo2009}
J.~Russo, P.~Tartaglia and F.~Sciortino, \emph{J. Chem. Phys.}, 2009,
  \textbf{131}, 014504\relax
\mciteBstWouldAddEndPuncttrue
\mciteSetBstMidEndSepPunct{\mcitedefaultmidpunct}
{\mcitedefaultendpunct}{\mcitedefaultseppunct}\relax
\EndOfBibitem
\bibitem[SantaLucia(1996)]{SantaLucia1996}
J.~SantaLucia, \emph{Nucleic Acid Res.}, 1996, \textbf{28}, 1929\relax
\mciteBstWouldAddEndPuncttrue
\mciteSetBstMidEndSepPunct{\mcitedefaultmidpunct}
{\mcitedefaultendpunct}{\mcitedefaultseppunct}\relax
\EndOfBibitem
\bibitem[Rubinstein and Colby(2003)]{Rubinstein2003}
M.~Rubinstein and R.~Colby, \emph{Polymer Physics}, OUP Oxford, 2003\relax
\mciteBstWouldAddEndPuncttrue
\mciteSetBstMidEndSepPunct{\mcitedefaultmidpunct}
{\mcitedefaultendpunct}{\mcitedefaultseppunct}\relax
\EndOfBibitem
\bibitem[Sciortino(2019)]{Sciortino2019}
F.~Sciortino, \emph{Riv. del Nuovo Cim.}, 2019, \textbf{42}, 511--548\relax
\mciteBstWouldAddEndPuncttrue
\mciteSetBstMidEndSepPunct{\mcitedefaultmidpunct}
{\mcitedefaultendpunct}{\mcitedefaultseppunct}\relax
\EndOfBibitem
\bibitem[Corezzi \emph{et~al.}(2012)Corezzi, Fioretto, and
  Sciortino]{Corezzi2012}
S.~Corezzi, D.~Fioretto and F.~Sciortino, \emph{Soft Matter}, 2012, \textbf{8},
  11207\relax
\mciteBstWouldAddEndPuncttrue
\mciteSetBstMidEndSepPunct{\mcitedefaultmidpunct}
{\mcitedefaultendpunct}{\mcitedefaultseppunct}\relax
\EndOfBibitem
\bibitem[Hill(1987)]{Hill1987}
T.~Hill, \emph{Statistical Mechanics: Principles and Selected Applications},
  Dover Publications, 1987\relax
\mciteBstWouldAddEndPuncttrue
\mciteSetBstMidEndSepPunct{\mcitedefaultmidpunct}
{\mcitedefaultendpunct}{\mcitedefaultseppunct}\relax
\EndOfBibitem
\bibitem[Markham and Zuker(2005)]{Markham2005}
N.~Markham and M.~Zuker, \emph{Nucleic Acid Res.}, 2005, \textbf{33},
  W577\relax
\mciteBstWouldAddEndPuncttrue
\mciteSetBstMidEndSepPunct{\mcitedefaultmidpunct}
{\mcitedefaultendpunct}{\mcitedefaultseppunct}\relax
\EndOfBibitem
\bibitem[Smallenburg and Sciortino(2013)]{Smallenburg2013}
F.~Smallenburg and F.~Sciortino, \emph{Nat. Phys.}, 2013, \textbf{9},
  554--558\relax
\mciteBstWouldAddEndPuncttrue
\mciteSetBstMidEndSepPunct{\mcitedefaultmidpunct}
{\mcitedefaultendpunct}{\mcitedefaultseppunct}\relax
\EndOfBibitem
\bibitem[Rovigatti \emph{et~al.}(2019)Rovigatti, Gnan, Ninarello, and
  Zaccarelli]{Rovigatti2019}
L.~Rovigatti, N.~Gnan, A.~Ninarello and E.~Zaccarelli, \emph{Macromolecules},
  2019, \textbf{52}, 4895\relax
\mciteBstWouldAddEndPuncttrue
\mciteSetBstMidEndSepPunct{\mcitedefaultmidpunct}
{\mcitedefaultendpunct}{\mcitedefaultseppunct}\relax
\EndOfBibitem
\bibitem[Hubbard and Douglas(1993)]{Hubbard1993}
J.~Hubbard and J.~Douglas, \emph{Phys. Rev. E}, 1993, \textbf{47}, R2983\relax
\mciteBstWouldAddEndPuncttrue
\mciteSetBstMidEndSepPunct{\mcitedefaultmidpunct}
{\mcitedefaultendpunct}{\mcitedefaultseppunct}\relax
\EndOfBibitem
\end{mcitethebibliography}
\bibliographystyle{rsc} 

\newpage
\pagestyle{empty}


\section*{Supplementary material}

Supplementary Material to *Hyperbranched DNA clusters*
E. Lattuada, D. Caprara, V. Lamberti, and F. Sciortino

## A. Nupack DNA analysis

To confirm the thermodynamic binding behavior of the designed sequences, we use the NUPACK oligo simulator [1]. Based on SantaLucia thermodynamics calculations, NUPACK provides the melting profile for arbitrary sequences of oligomers at desired concentrations and salt conditions (more precisely, the $T$-dependence of the fraction of unbonded base pairs).

Fig. S1, shows the $T$-behavior associated with the self-assembly of the tetravalent particles as well as of the isolated sticky sequences (forming the $AB$ bonds). Here, the strand concentrations are fixed at $313\,\mu$M (corresponding to the largest sample concentration $c = 20\,\mathrm{mg/ml}$) and the salt concentration at $250\,\mathrm{mM}$ of NaCl, respectively. As can be noted, the gap in the melting temperatures between the particle assembly (see magenta points) with respect to their sticky-end hybridization (blue) guarantees a net separation between the self-assembly of the nanostructures ($T_{\mathrm{NS}} \simeq 77°$C) and the formation of the interparticle bonds ($T_b \simeq 42°$C).

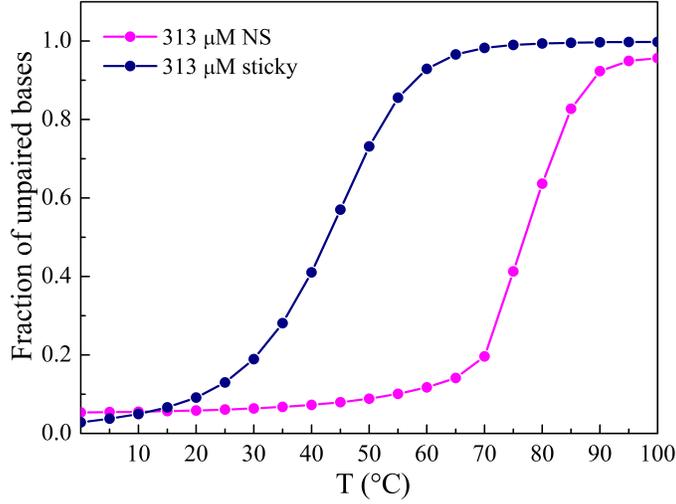

Figure S1: Melting curves calculated using NUPACK oligo analyzer [1] for both the NS arms (magenta) and the sticky tips (blue). Note that the self-assembly of the NS precedes on cooling the binding of the sticky sequences.

## B. Considerations about the cluster free-energy and Flory-Stockmayer approach

Here, we relate the bond probability assumed in the Flory-Stokmayer (FS) theory [2, 3] to the partition function of a bond. The FS theory assumes a value for the bond probability $p_b$. In the hyperbranched $AB_{f-1}$ case, $p_b = \#_b/N$, being $\#_b$ the number of bonds in the system and $N$ the total number of possible bonds (equal to the number of monomers). In a thermodynamic approach, $p_b$ is controlled by the interaction potential and by the state variables $T$ and $V$.

Let us consider a solution of $N$ $A$-patches and $N(f-1)$ $B$-patches. We assume that an equilibrium is established between bonded and non-bonded interactions, the latter quantified by a bonding volume $V_b$ and a bonding energy $\epsilon_{AB}$. The mass-action law relates the number of reacted ($N_{AB}$) and unreacted ($N_A$ and $N_B$) $A$- and $B$-patches to their partition functions as [4]

$$\frac{N_{AB}}{N_A N_B} = \frac{Q_{AB}}{Q_A Q_B}. \tag{1}$$

Neglecting intramolecular effects, to a first approximation, $Q_A = Q_B = V/\Lambda^3$ (where $\Lambda$ is the de Broglie wavelength originating from the integration over momenta) and

$$Q_{AB} = \frac{V V_b}{\Lambda^6} e^{-\beta \epsilon_{AB}}, \tag{2}$$

so that

$$\frac{N_{AB}}{N_A N_B} = \frac{V_b}{V} e^{-\beta \epsilon_{AB}}. \tag{3}$$

Since $N_{AB}$ is equal to the number of $AB$ bonds in the system ($\#_b$), we can write for the number of unreacted $A$- and $B$-patches, respectively, $N_A = N - \#_b$ and $N_B = N(f-1) - \#_b$. Hence, the left hand side of Eq. (1) can be rewritten as

$$\frac{N_{AB}}{N_A N_B} = \frac{\#_b}{(N - \#_b)[N(f-1) - \#_b]}$$
$$= \frac{p_b}{(1 - p_b)[(f-1) - p_b]}, \tag{4}$$

from which it follows that

$$\frac{p_b}{(1 - p_b)[(f-1) - p_b]} = \frac{N V_b}{V} e^{-\beta \epsilon_{AB}}. \tag{5}$$

This last relation provides the connection between the model parameters ($V_b, \epsilon_{AB}$), $V$, and $T$ and the bond probability $p_b$.

Starting from the thermodynamics expression for non-interacting clusters, one can write the number of clusters of size $n$ resulting from the association process of the $A$- and $B$-patches as

$$N_c(n) = Q_n z^n, \tag{6}$$

where $z$ plays the role of the Lagrange multiplier controlling the total number of particles in the system, which can also be expressed in terms of the number of unreacted particles (clusters of size 1, $N_c(1)$), since

$$z = \frac{N_c(1)}{Q_1} = \frac{N_c(1) \Lambda^3}{V}. \tag{7}$$

The number of free monomers can be estimated using the FS relation

$$N_c(1) = N(1 - p_b) \left(1 - \frac{p_b}{f - 1}\right)^{f-1}, \tag{8}$$

which expresses the fact that all the monomers must have one $A$ and $(f-1)$ $B$ sites unbonded (we recall that the probability that a $B$ site is unbonded is $\#_b/[N(f-1)]$). Then,

$$\frac{N_c(1)}{V} = \rho(1 - p_b) \left(1 - \frac{p_b}{f - 1}\right)^{f-1} \tag{9}$$

and we can write, by defining $\rho = N/V$ and $\rho_1 = N_c(1)/V$, and equating Eq. (7) and Eq. (9),

$$\frac{\rho_1}{\rho} = \frac{z \Lambda^{-3}}{\rho} = (1 - p_b) \left(1 - \frac{p_b}{f - 1}\right)^{f-1}. \tag{10}$$

Substituting this expression (to the power $n$) in the FS cluster size distribution (Eq. (1) of the manuscript, here reproduced)

$$N_c(n) = N(1 - p_b) \frac{[(f-1)n]!}{n![(f-2)n+1]!} \frac{p_b^{n-1}(f - 1 - p_b)^{(f-2)n+1}}{(f-1)^{(f-1)n}}, \tag{11}$$

one finds

$$N_c(n) = N(1 - p_b) D(n, f) p_b^{n-1} \frac{(f - 1 - p_b)^{1-n}}{(1 - p_b)^n} \left(\frac{z \Lambda^{-3}}{\rho}\right)^n, \tag{12}$$

where

$$D(n, f) = \frac{[(f-1)n]!}{n![(f-2)n+1]!}. \tag{13}$$

Simplifying and making use of the relation in Eq. (5),

$$\frac{N_c(n)}{V} = \frac{N}{V} D(n,f) \left(\frac{NV_b}{V} e^{-\beta \epsilon_{AB}}\right)^{n-1} \left(\frac{z\Lambda^{-3}}{\rho}\right)^n$$

$$= D(n,f) \left(V_b e^{-\beta \epsilon_{AB}}\right)^{n-1} (z\Lambda^{-3})^n. \quad (14)$$

Defining a reference volume $V_{\text{ref}}$, one can define a reference bonding free-energy $\mathcal{F}_{\text{bond}}$ as [5]

$$e^{-\beta \mathcal{F}_{\text{bond}}(V_{\text{ref}},T)} = \frac{V_b}{V_{\text{ref}}} e^{-\beta \epsilon_{AB}}, \quad (15)$$

such that

$$N_c(n) = \frac{V}{V_{\text{ref}}} D(n,f) \left(e^{-\beta \mathcal{F}_{\text{bond}}(V_{\text{ref}},T)}\right)^{n-1} (zV_{\text{ref}}\Lambda^{-3})^n. \quad (16)$$

By comparing the definition of $\mathcal{F}_{\text{bond}}$ and Eq. (5), it follows that

$$\frac{p_b}{(1-p_b)[(f-1)-p_b]} = \frac{NV_{\text{ref}}}{V} e^{-\beta \mathcal{F}_{\text{bond}}}, \quad (17)$$

which provides the link between the bond probability and the bonding free-energy.

Redefining $z$ as $(zV_{\text{ref}}\Lambda^{-3})$ and comparing Eq. (16) and Eq. (6), the partition function of a cluster of size $n$ with no loops (NL) can be identified as

$$Q_n^{\text{NL}} = \frac{V}{V_{\text{ref}}} D(n,f) \left(e^{-\beta \mathcal{F}_{\text{bond}}(V_{\text{ref}},T)}\right)^{n-1}, \quad (18)$$

which clearly shows the cluster center of mass contribution $V$, the combinatorial contribution $D(n,f)$, and the contribution arising from the $n-1$ bonds.

Assuming that loops are also possible, one needs to add the contribution which includes the loops to the FS partition function. The additional intracluster bond adds a term proportional to $V_b e^{-\beta \epsilon_{AB}}$. Then, we propose to write the partition function of clusters with loops (L) as

$$Q_n^{\text{L}} = Q_n^{\text{NL}} g(n,\beta) e^{-\beta \mathcal{F}_{\text{bond}}(V_{\text{ref}},T)}, \quad (19)$$

where the system-dependent factor $g(n,\beta)$ accounts for the additional free-energy change. Specifically, it includes any combinatorial and any elastic free-energy contributions not accounted by the $V_b e^{-\beta \epsilon_{AB}}$ term. Hence, the resulting cluster size distribution is

$$N_c(n) = N_{\text{NL}}(n) + N_{\text{L}}(n) \quad (20)$$

$$= \frac{V}{V_{\text{ref}}} D(n,f) \left(e^{-\beta \mathcal{F}_{\text{bond}}(V_{\text{ref}},T)}\right)^{n-1} \left(1 + g(n,\beta) e^{-\beta \mathcal{F}_{\text{bond}}(V_{\text{ref}},T)}\right) z^n,$$

where $g(n,\beta)$ can be evaluated by calculating the ratio between the number of clusters of size $n$ with and without loops, $N_{\text{L}}(n)/N_{\text{NL}}(n)$.

Note that, at low $T$, when all bonds are formed, $g(n,\beta)e^{-\beta \mathcal{F}_{\text{bond}}(V_{\text{ref}},T)} \gg 1$ and only the contribution from the clusters with a loop to the partition function survives and

$$Q_n = \frac{V}{V_{\text{ref}}} D(n,f) \left(e^{-\beta \mathcal{F}_{\text{bond}}(V_{\text{ref}},T)}\right)^n g(n,\beta) \quad (21)$$

and

$$N_c(n) = \frac{V}{V_{\text{ref}}} D(n,f) g(n,\beta) \left(z e^{-\beta \mathcal{F}_{\text{bond}}(V_{\text{ref}},T)}\right)^n. \quad (22)$$

This time

$$N_c(1) = \frac{V}{V_{\text{ref}}} g(1,\beta) \left( z e^{-\beta \mathcal{F}_{\text{bond}}(V_{\text{ref}},T)} \right), \tag{23}$$

so that

$$z e^{-\beta \mathcal{F}_{\text{bond}}(V_{\text{ref}},T)} = \frac{\rho_1}{g(1,\beta)}, \tag{24}$$

where $\rho_1$ is the monomer's (unbonded particles) density when the volume is measured in units of $V_{\text{ref}}$, and

$$N_c(n) = D(n,f) \, g(n,\beta) \left( \frac{\rho_1}{g(1,\beta)} \right)^n, \tag{25}$$

which, neglecting the weak dependence on $T$ of $g(1,\beta)$, does not depend on $T$ any longer. The cluster size distribution has reached its "ground state" limit.

### C. DLS data fitting

Fig. S2 compares the experimental data for the autocorrelation functions $g_1(t)$ with the fit performed using the double stretched exponential function of Eq. (13) in the main document at high (a), intermediate (b), and low temperatures (c).

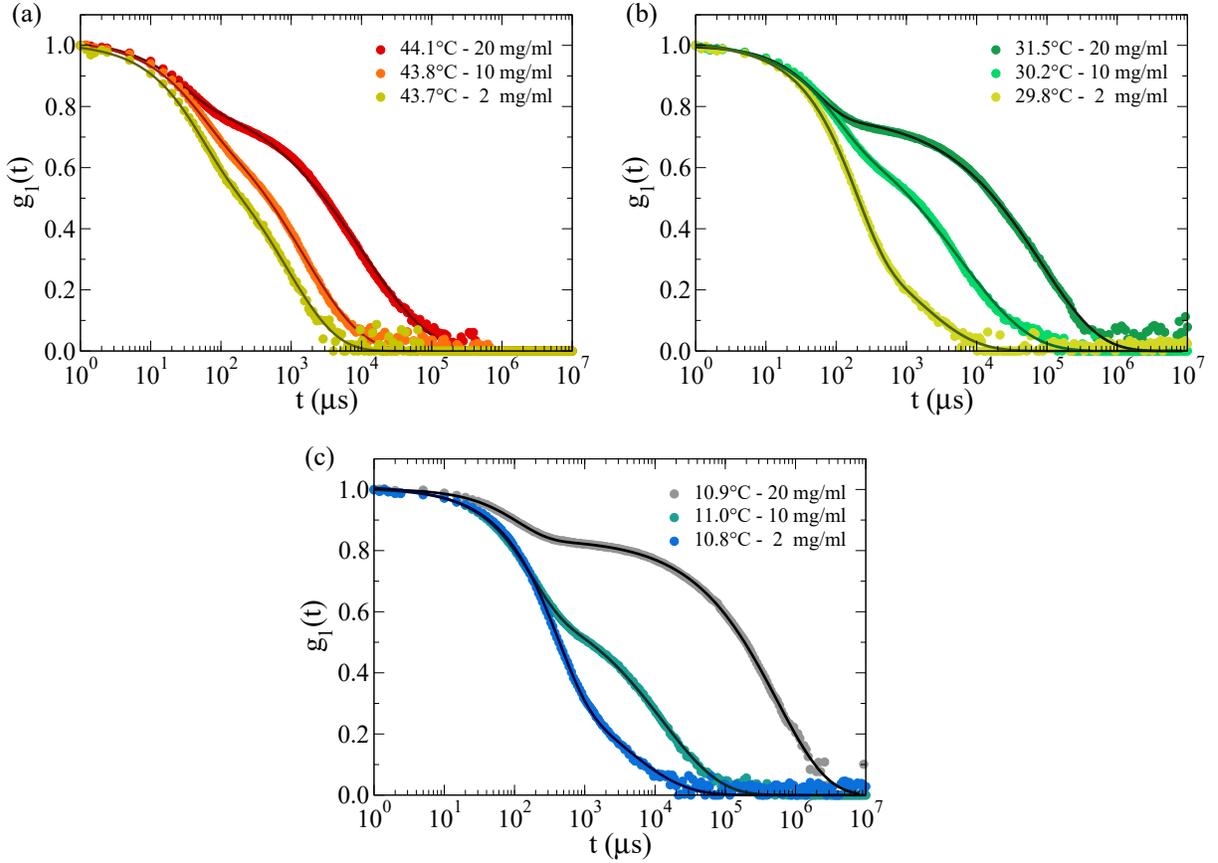

Figure S2: Comparison of the experimental data (symbols) and the fit with Eq. (13) in the main document (full lines) for the autocorrelation functions $g_1(t)$ obtained for the three different investigated NS concentrations at high (a), intermediate (b), and low temperatures (c).

### D. Hydrodynamic radius from simulations

In Fig. S3, we show the hydrodynamic radius calculated from the simulations for $c = 2\,\mathrm{mg/ml}$ and $c = 20\,\mathrm{mg/ml}$ according to the method described in Ref. [6]. Briefly, for each cluster in the system, we computed the smallest convex set of points (the convex hull) that encloses the position of the bases of the NSs forming the clusters. Then, the gyration tensor is calculated using the center of mass of the triangles composing the convex hull. Finally, the gyration tensor is diagonalized, obtaining the three eigenvalues $\lambda_1$, $\lambda_2$, and $\lambda_3$, which are related to the three semi-axes $\{a\}$ of the ellipsoid as $a_i = \sqrt{3\lambda_i}$. The hydrodynamic radius is evaluated according to

$$R_h = \frac{2}{\int_0^\infty \left[(a_1^2 + \theta)(a_2^2 + \theta)(a_3^2 + \theta)\right]^{-\frac{1}{2}} \, d\theta}. \tag{26}$$

As done for the radius of gyration in the main text, we calculated the ensemble average of the hydrodynamic radius over clusters of same size $n$ and different simulation times.

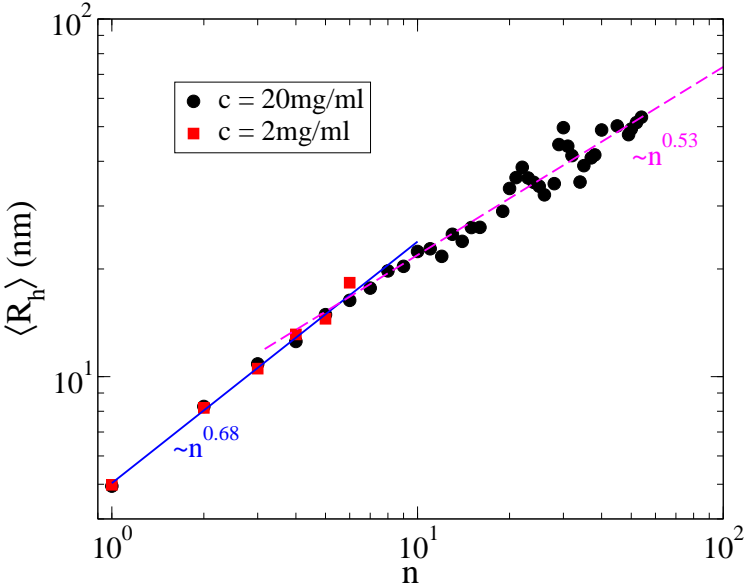

Figure S3: Hydrodynamic radii calculated from the simulations for $c = 2\,\mathrm{mg/ml}$ (black points) and $c = 20\,\mathrm{mg/ml}$ (red squares), respectively. Blue solid line and magenta dashed line are the power law fits to the data for $n \leq 5$ and $n > 5$, respectively.

[1] J. Zadeh, C. Steenberg, J. Bois, B. Wolfe, M. Pierce, A. Khan, R. Dirks and N. Pierce, *J. Comput. Chem.*, 2011, **32**, 170–173.
[2] P. Flory, *J. Am. Chem. Soc.*, 1941, **63**, 3083.
[3] M. Rubinstein and R. Colby, *Polymer Physics*, OUP Oxford, 2003.
[4] T. Hill, *Statistical Mechanics: Principles and Selected Applications*, Dover Publications, 1987.
[5] F. Sciortino, E. Bianchi, J. Douglas and P. Tartaglia, *J. Chem. Phys.*, 2007, **126**, 194903.
[6] L. Rovigatti, N. Gnan, A. Ninarello and E. Zaccarelli, *Macromolecules*, 2019, **52**, 4895.



\end{document}